\documentclass[12pt,preprint]{aastex}

\usepackage{ifpdf}
\usepackage{rotating}
\usepackage{lscape}
\usepackage{mdwlist}

\def \msun   {\hbox{M$_\odot$}}

\shorttitle{Northern \emph{NYMG} Candidates}
\shortauthors{Schlieder et al.}

\begin{document}

\title{Cool Young Stars in the Northern Hemisphere:  $\beta$ Pictoris and AB Doradus Moving Group Candidates}

\author{Joshua E. Schlieder\altaffilmark{1,2,3}, S\'{e}bastien L\'{e}pine\altaffilmark{4}, and Michal Simon\altaffilmark{1,3}}
\altaffiltext{1}{Department of Physics and Astronomy, Stony Brook University,
    Stony Brook, NY 11794, michal.simon@stonybrook.edu}
\altaffiltext{2}{Currently at Max-Planck-Institut f\"{u}r Astronomie, K\"{o}nigstuhl 17, 69117 Heidelberg, Germany, schlieder@mpia-hd.mpg.de}
\altaffiltext{3}{Visiting Astronomer, NASA Infrared Telescope Facility (IRTF).}
\altaffiltext{4}{Department of Astrophysics,  American Museum of Natural History, Central Park West at 79th Street, New York, NY 10024, lepine@amnh.org}

\author{(Accepted to the Astronomical Journal)}

\begin{abstract}
As part of our continuing effort to identify new, low-mass members of nearby, young moving groups (\emph{NYMGs}), we present a list of young, low-mass candidates in the northern hemisphere.  We used our proven proper motion selection procedure and \emph{ROSAT}-X-ray and \emph{GALEX}-UV activity indicators to identify 204 young stars as candidate members of the $\beta$ Pictoris and AB Doradus \emph{NYMGs}.  Definitive membership assignment of a given candidate will require a measurement of its radial velocity and distance.  We present a simple system of indices to characterize the young candidates and help prioritize follow up observations.  New group members identified in this candidate list will be high priority targets for: 1) exoplanet direct imaging searches, 2) the study of post-T-Tauri astrophysics, 3) understanding recent locl star formation, and 4) the study of local galactic kinematics.  Information available now allows us to identify 8 likely new members in the list.  Two of these, a late-K and an early-M dwarf, we find to be likely members of the $\beta$ Pic group. The other six stars are likely members of the AB Dor moving group.  These include an M dwarf triple system, and three very cool objects that may be young brown dwarfs, making them the lowest-mass, isolated objects proposed in the AB Dor moving group to date.   
\end{abstract}

\section{Introduction}

Well characterized samples of low-mass, pre-main sequence (\emph{PMS}) stars are important for understanding star formation and evolution, the circumstellar environment, and planetary system formation.   These stars are attractive targets for exoplanet searches by direct imaging because young, massive planets are expected to be self-luminous from gravitational contraction and the contrast between host star and planet is more favorable when the star is intrinsically faint.  The nearest and youngest stars in clusters lie in the Taurus and Ophiuchus star forming regions (\emph{SFRs}) with ages $\sim$1 - 5 Myr and  distances 120 - 145 pc.  Their distances limit the detail with which they can be studied.  Nearby, low-mass \emph{PMS} stars that are no longer associated with their \emph{SFRs} have long been sought following Herbig's (1978) suggestion that they may exist in large numbers mingled with field stars.  Post-T-Tauri stars (\emph{PTTS}), with ages $\sim$10$^7$ --  $\sim$10$^8$ yr, exhibit properties associated with stellar youth, such as chromospheric activity, lithium absorption, and rapid rotation, which provide the means to distinguish them among older field stars.    

Stellar counterparts to \emph{ROSAT} (Voges et al. 1999, 2000; hereafter V99, V00) X-ray sources and the compilation of precision astrometric catalogs (e.g. the \emph{Hipparcos} catalog, Perryman et al. 1997) have yielded the identification of nearby, young stars as members of coeval groups characterized by the common motion of their members through the Galaxy.  These groups of stars are known as nearby, young moving groups (\emph{NYMGs}) (see Zuckerman \& Song 2004, hereafter ZS04; Torres et al. 2008, hereafter T08).  Members of \emph{NYMGs} have ages $<$100 Myr and lie at distances $\lesssim$100 pc (T08).  The ages, space motions, and sky distributions of stars in the \emph{NYMGs} suggest that they share a common origin.  Kinematic traceback studies show they may be related to a star formation event in the Sco-Cen region (Mamajek \& Feigelson 2001, Fern\'andez et al. 2008). 

Most members of the known moving groups are concentrated in the south.  However, we began our search for new, low-mass \emph{NYMG} members in the $\beta$ Pictoris moving group (\emph{BPMG}, L\'epine \& Simon 2009, hereafter LS09) and expanded it to the AB Doradus moving group (\emph{ABDMG}, Schlieder et al. 2010, hereafter S10) because they have known members in the north.  The northern hemisphere presents a relatively untapped resource in the search for moving group members. 

To date, confirmed members of the \emph{BPMG} and \emph{ABDMG} show a marked deficiency in low-mass stars between spectral subtypes M0 to M6.  If those moving groups have a mass function consistent with field stars (see Bochanski et al. 2010), then astronomers are currently missing the bulk of the $\beta$ Pic and AB Dor members.  There are two primary factors contributing to the observed deficiency of known low-mass members: 1) low-mass stars are intrinsically faint, thus the majority are beyond the magnitude limits of the \emph{Hipparcos} and \emph{Tycho-2} astrometric catalogs which, along with the \emph{ROSAT} catalogs, have so far been used as the primary sources for identifying new group members.  2) The lowest-mass stars lack reliable youth diagnostics from which the missing stars could be easily identified.  Traditional activity indicators, X-ray and H$\alpha$ emission, are increasingly less reliable for spectral types (\emph{SpTy}) later than M4 because most older, field stars are active in that spectral range (West et al. 2008, 2011, hereafter W11).  This is probably a consequence of the transition to fully convective interiors.  Lithium is also depleted for most low-mass stars at the ages of the \emph{NYMGs} (Palla \& Randich 2004) which makes the detection/non-detection of the Li 6708~\AA~line an unreliable indicator of age.  Furthermore, Baraffe and Chabrier (2010, hereafter BC10) have also shown that Li depletion may be strongly affected by stellar accretion history in low-mass stars.

LS09 developed an astrometric technique to identify low-mass candidates of \emph{NYMGs} in proper motion catalogs. The projected mean motion vector of a known moving group is used to identify candidates based on their proper motion and optical/IR photometry.  The visual magnitude of the proper motion samples to which LS09 applied this technique ($V<12$) limited the identification of likely new members to stars earlier than approximately M2 \emph{SpTy} (LS09).  S10 expanded the technique to a preliminary version of the deeper ($V<19$) \emph{SUPERBLINK} catalog (\emph{SBK}, L\'epine et al. 2012, in prep.) to probe into the mid-M range.

Several other efforts are underway to identify low-mass \emph{NYMG} members.  Shkolnik et al. (2009, hereafter Sh09) use \emph{ROSAT} X-ray data and spectroscopic follow up to identify nearly 150 nearby, young M dwarfs in a d $<$ 25 pc sample, at least some of which are likely to be \emph{NYMG} members.  Shkolnik et al. (2011, hereafter Sh11) present an analysis of M dwarfs in the \emph{HST} Guide Star Catalogue (Lasker et al. 2008) and the \emph{2MASS} All-Sky Catalog of Point Sources (\emph{2MASS}, Skrutskie et al. 2006) having \emph{GALEX} counterparts to calibrate UV emission as a youth indicator in low-mass stars.  UV emission was then used as a basis to identify two new members of the $\sim$10 Myr old TW Hydrae association.  Kiss et al. (2011) use a selection technique similar to T08 and data from The Radial Velocity Experiment (\emph{RAVE}, Steinmetz et al. 2006) to identify new \emph{NYMG} members in the south.  Rice et al. (2010) report the first isolated brown dwarf member of the \emph{BPMG} as a result of a large scale astrometric and spectroscopic survey of brown dwarfs.  Others have started to use \emph{GALEX} data to identify cool young stars as well, focusing on UV excess as a diagnostic (Findeisen \& Hillenbrand 2010, hereafter FH10; Rodriguez et al. 2011, hereafter R11).  A program is also underway that uses a selection technique very similar to LS09, S10, and this work but assigns membership probabilities to candidates using Bayesian analysis techniques.  This search focuses on $\beta$ Pic, AB Dor, and Tuc/Hor candidates (see Rice et al. 2011).

In \S 2 and \S3 of this paper we apply the techniques described in LS09 and S10 to search for \emph{BPMG} and \emph{ABDMG} candidates in the \emph{Tycho-2} catalog (H{\o}g et al. 2000, hereafter H00), the \emph{LSPM-North} catalog (L\'epine and Shara 2005, herafter LS05), and the now complete northern hemisphere \emph{SBK} catalog.  This extension permits identification of moving group candidates with \emph{SpTy's} beyond M4, which raises the problem of identifying the young stars among them.  For mid-M dwarfs and later, the gravity sensitive alkali lines, such as the neutral sodium lines (\emph{NaI}), can serve as proxies for youth.  At a few tens of Myrs age, these stars are still contracting to the main sequence (\emph{MS}) and have lower photospheric gravities than they will have later; this makes the \emph{NaI} doublet at $\sim$8200~\AA~a useful indicator of age in low-mass stars (Schlieder et al. 2011).  The extensive spectroscopic observations required to study this feature are however beyond the scope of this paper.  We therefore describe in \S 4 our list of young $\beta$ Pic and AB Dor candidates in the north identified using our selection technique and the X-ray/UV youth criteria described in Sh09 and Sh11.  We also carry out a preliminary study of the candidates to determine those for which these youth indicators are most reliable.  In \S5 we investigate the bulk properties of the sample to guide follow up priority.  In  \S 6 we describe likely new members identified in the analysis presented and we summarize our results in \S 7.

\section{Candidate Selection}

T08 lists 50 \emph{BPMG} members with membership probability $\ge$90\% for all but 3 stars\footnote{Proposed \emph{BPMG} members HD 203, HD 15115, and HD 199143 have probabilities 75\%, 60\%, and 75\% respectively}.  The membership probabilities were calculated using the full, six-dimensional galactic kinematics of the group in a \emph{k-NN} model (T08).  We include all 50 known members in our analysis.  The known \emph{BPMG} members lie at a median distance of $\sim$35 pc and have ages 10-20 Myr (T08).  The motion of the group through the galaxy is defined using \emph{U}, \emph{V}, and \emph{W} space velocities (Johnson and Soderblom 1987)  with means (\emph{U$_{BPMG}$}, \emph{V$_{BPMG}$}, \emph{W$_{BPMG}$}) = (-10.1$\pm$2.1, -15.9$\pm$0.8, -9.2$\pm$1.0) km s$^{-1}$ relative to the Sun (T08).  In this coordinate system, \emph{U} is positive toward the galactic center, \emph{V} is positive in the direction of solar motion around the Galaxy, and \emph{W} is positive toward the north galactic pole.  The group shows extension in the direction toward the galactic center (Fig.~\ref{bp_xyz}), a feature common to all \emph{NYMGs} younger than 30 Myr  (T08)\footnote{\emph{XYZ} distances are defined positive in the same directions as \emph{UVW} velocities.}.

The 89 known members of the \emph{ABDMG} lie at a median distance of $\sim$30 pc and have ages of $\sim$70 Myr (T08).  All of these stars have a membership probability in T08 $\ge$85\%.  The mean velocities of AB Dor group members are (\emph{U$_{ABDMG}$}, \emph{V$_{ABDMG}$}, \emph{W$_{ABDMG}$}) = (-6.8$\pm$1.3, -27.2$\pm$1.2, -13.3$\pm$1.6) km s$^{-1}$.  These space velocities are comparable to those of Pleiades open cluster members.   Luhman et al. (2005) and Ortega et al. (2007) argue that the \emph{ABDMG} may be remnant of the star formation event that formed the Pleiades.  Possibly coincidentally, the space velocities of some AB Dor stars are similar to $\beta$ Pic stars in the \emph{U} and \emph{W} plane.  The group lacks the \emph{X} direction extension of younger \emph{NYMGs} and exhibits a more uniform galactic distance distribution (see Fig.~\ref{abd_xyz}). 

LS09 and S10 have described the proper motion selection algorithm in detail.  To produce a list of young candidates in the northern hemisphere we performed only the first 3 steps in the search procedure: 1) isolate a sample of stars in a proper motion catalog whose proper motion vectors are consistent with moving group membership (see LS09 Eqns. (1-4)), 2) identify stars in that subsample whose photometric distance (\emph{d$_{phot}$}) is consistent with the kinematic distance (\emph{d$_{kin}$}), which is derived from the proper motion assuming assuming group membership (see LS09 Eqns. (5-7)), 3) trim this sample to include only stars exhibiting indicators of youth.  After this, all that will be required is a confirmation that the radial velocity (\emph{RV}) and astrometric distance are consistent with \emph{NYMG}
membership. This follow up work is now in the planning stages.  We follow the labeling convention introduced in S10:

\begin{enumerate}
\item
{\emph{Candidate} - A low-mass star having proper motion and photometry consistent with \emph{NYMG} membership} 
\item
{\emph{Probable Young Candidate (PYC)} - A candidate exhibiting indicators of youth}
\item
{\emph{Likely New Member (LNM)} - A \emph{PYC} having a \emph{RV} or distance consistent with \emph{NYMG} membership}
\end{enumerate}

Because the search parameters govern the candidates selected, we describe the parameters in detail and define the limits used in this search:

\noindent{\emph{Lower Proper Motion Limit}, $\mu_{min}$:  A lower limit that is very small will introduce contamination 
from field stars whose proper motions align with the projected mean motion of the group by chance. We chose 
$\mu_{min}$~=~40 mas yr$^{-1}$ as a lower limit because the majority of known \emph{BPMG} and \emph{ABDMG} members have 40 mas yr$^{-1}$ $\le$ $\mu$ $\le$ 200 mas yr$^{-1}$  (see Fig.~\ref{pm_plot}).}

\noindent{\emph{Dispersion About Average Space Motion, $\phi$}:  The scalar product of the mean projected motion of the group with the proper motion of a catalog star, calculated in the plane of the sky local to the star, defines cos$\phi$ (LS09, Eqn. 3).  Stars that are actual \emph{NYMG} members will have $\phi$ close to 0.  The largest acceptable value of $\phi$, called $\phi_{max}$, depends on the \emph{UVW} velocity dispersion of known moving group members and can be assessed by calculating $\phi$ for each known member (see Fig.~\ref{phi_plot}).   We choose $\phi_{max}$~=~10$^{\circ}$ because it includes most of the known \emph{BPMG} and \emph{ABDMG} members and limits contamination from kinematic interlopers.}

\noindent{\emph{Kinematic Distance, d$_{kin}$}:  If one assumes that a candidate selected by its proper motion actually belongs to the moving group considered, then its \emph{d$_{kin}$} can be calculated from the magnitude of the proper vector (LS09, Eqn. 6) and used as a selection parameter in two ways.  First, a comparison of candidate and known member \emph{d$_{kin}$} can act as a selection cut.  The range of \emph{d$_{kin}$} accepted when searching for candidates of an \emph{NYMG} is again determined from the distribution of known members (Fig.~\ref{dkin_plot}).  Both \emph{d$_{kin}$} distributions peak between 30-40 pc with an upturn at 70 pc.  Three of the 5 stars in the 70-80 pc range in the \emph{BPMG} are later than \emph{SpTy} K4, and only one has a measured parallax.  The upturn at 70 pc in the \emph{ABDMG} is more drastic, with more than 3 times the stars in the previous bin.  All 15 of the proposed \emph{ABDMG} members in the 70-80 pc bin are earlier than mid-K type.  However, only 2 have measured parallaxes.  Since most of the stars in the 70-80 pc bins and beyond do not have measured distances we conservatively choose \emph{d$_{kin}$}~$\le$~70 pc as the cutoff in the candidate search.}    

\noindent{Second, the \emph{d$_{kin}$} is used to calculate a pseudo-absolute \emph{K} magnitude (\emph{M$_k$}) in an \emph{M$_k$} \emph{vs.} (\emph{V-K$_s$}) color-magnitude diagram (\emph{CMD}).  Candidates that are true \emph{NYMG} members will be positionally and photometrically consistent with the cluster sequence of known group members in the \emph{CMD} (see Fig.~\ref{cmd_plot}).  Any candidate remaining in the sample after previous cuts that falls outside of the cluster-sequence locus is presumed to have its true distance over or under-estimated by \emph{d$_{kin}$} and is rejected from the sample.


\noindent{\emph{(\emph{V-K$_{s}$}) Color}:  A lower limit on the (\emph{V-K$_{s}$}) color determines the upper bound on the mass of the candidates.  A 0.7 $M_{\odot}$ dwarf with an age of 40 Myr will be SpTy $\sim$K7 and have (\emph{V-K$_{s}$})~$\approx$~3.2 (Siess et al. 2000, hereafter SDF2000).  We used this value to concentrate our search on M dwarf candidates and avoid a high level of giant contamination (see Fig.~\ref{cmd_plot}). }

Proper motion catalogs are magnitude and proper motion limited. Thus, searching several catalogs with complementary limits allows for the identification of all candidates within the combined limits of the search parameters and catalogs.  The combined properties of the catalog subsamples we searched for northern hemisphere candidates are $\delta$ $\ge$ 0$^{\circ}$, $\mu$ $\ge$ 40 mas yr$^{-1}$, and complete to \emph{V} = 19 mag.  The individual catalogs we searched are as follows:

\noindent{{\it The Tycho-2} catalog contains positions, proper motions, and photometric data for $\sim$2.5 million stars.  The catalog is $\sim$99\% complete to \emph{V}~$\approx$~11.0, with a limiting magnitude of \emph{V}~$\approx$~13.0, and has proper motion accuracy of 2.5 mas yr$^{-1}$ (H00).  We selected a subsample of northern hemisphere \emph{Tycho-2} stars with $\mu$ $\geq$ 40 mas yr$^{-1}$ ($\sim$120,000 stars) which is cross correlated with \emph{2MASS}  to obtain near-IR \emph{J}, \emph{H}, and \emph{K$_{s}$} magnitudes.  Our previous results show that searches in the \emph{Tycho-2} catalog return candidates that are mostly earlier than \emph{SpTy} $\sim$M3 (S10), a result of the \emph{V} magnitude limit of the catalog.}
 
 
 \noindent{{\it The LSPM-North}  catalog is an astrometric catalog of positions, proper motions, and multi-band photometry of $\sim$62,000 high proper motion stars produced by data mining the Digitized Sky Surveys using specially developed software (LS05).  The catalog contains stars with $\mu$ $>$ 150 mas yr$^{-1}$ ($\mu_{err}$ $\approx$ 8 mas yr$^{-1}$) north of the celestial equator and is $\sim$99\% complete for 12.0 $<$ \emph{V} $<$ 19.0 with a faint limit of \emph{V} = 21.0 (LS05).  \emph{LSPM-North} complements the magnitude range of \emph{Tycho-2}, overlapping it in some cases, and allows access to \emph{NYMG} candidates down to $\sim$0.1 $\msun$.}
 
\noindent{{\it SUPERBLINK (SBK)} is the smaller proper motion (40 mas yr$^{-1}$ $\leq$ $\mu$ $\leq$ 150 mas yr$^{-1}$) counterpart of the \emph{LSPM-North} catalog and was produced using the same technique and thus retains the same limits (S\'ebastien L\'epine, private communication).  This database contains $\sim$1.5 million stars with $\delta$ $>$ 0$^{\circ}$.  The lower proper motion limit of \emph{SBK} allows access to a larger spatial volume and hence more potential NYMG candidates unavailable in the \emph{Tycho-2} or \emph{LSPM-North} catalogs.

\section{$\beta$ Pic and AB Dor Group Candidates}

We apply the search algorithm to the proper motion catalogs using the search parameter limits discussed in \S 2 to identify candidates of the \emph{BPMG} and \emph{ABDMG}.  We emphasize that our technique is statistical in nature; it can identify candidates only within the limits chosen to characterize a group.  The resulting list of candidates includes recovered known members,  stars previously investigated in this project (LS09, S10), contaminants, and many new \emph{NYMG} candidates, some of which will eventually become \emph{LNMs}.  We describe here the subsamples in the initial candidate list before presenting the final young sample.

The search for \emph{BPMG} candidates returns 132 stars (see Fig.~\ref{cmd_plot}).   Only 2 previously known \emph{BPMG} members are recovered (Table 1).  We expected to recover 4 members based on their declinations and colors.  The 2 not recovered, BD+30 397B and HIP 23418, are missing from the catalogs searched and fall outside of the \emph{BPMG} sequence locus defined in Fig. \ref{cmd_plot} respectively.  The remaining known BPMG members in T08 are not recovered because they lie in the southern hemisphere or have (\emph{V-K$_s$}) colors that are too blue.  Nineteen of the candidates were already investigated (LS09, S10), we therefore  delete these from the present list, leaving 111 \emph{BPMG} candidates.

The search for \emph{ABDMG} candidates identifies 582 stars.  Compared to the \emph{BPMG} this is a much larger number of candidates and is a consequence of the older age of the \emph{ABDMG}.   For a given \emph{SpTy}, a member of the \emph{ABDMG} is fainter than one in the \emph{BPMG}, thus the cluster sequence of the \emph{ABDMG} lies lower in the \emph{CMD} and the search algorithm picks up more contaminants  (see Fig.~\ref{cmd_plot}).  The candidate search recovered 4 known \emph{ABDMG} members (Table 1).  The declinations and colors of known \emph{ABDMG} members allow for the recovery of 6 in the search.  The 2 not recovered are HD 21845B, a close companion lacking sufficient photometric data in the catalogs, and BD+01 2447, which lies outside of the \emph{ABDMG} sequence locus defined in Fig. \ref{cmd_plot}.  S10's search included 9 of the candidates, we remove them from the sample leaving 569 \emph{ABDMG} candidates.

After removal of known members and previously investigated stars, the candidate searches in the \emph{BPMG} and \emph{ABDMG} result in a sample of 680 low-mass moving group candidates. These remaining stars represent a mixture of different populations including old and young dwarfs and evolved stars such as late type giants.  We may identify some of the interlopers by considering their locations in an ((\emph{J-H}) vs. (\emph{H-K$_{s}$})) color-color diagram.  Figure~\ref{ccd_plot} shows the candidate diagram.  We overplotted the expected sequences for A0 to M6 dwarfs and K to M giants (Bessel \& Brett 1988; Bessell 1991).   Nearly all of the candidates occupy the region expected for late-type dwarfs.  A few fall outside of this region; 2 align with the expected giant sequence at (\emph{J-H})$\sim$0.8 and fewer than 10 are coincident with the expected sequence for earlier dwarfs ((\emph{J-H}) $\lesssim$ 0.55 and (\emph{H-K$_{s}$}) $\approx$ 0.1).  These candidates are not included in further analyses.   The 7 candidates in the solid black box have colors representative of ultracool objects at the hydrogen burning limit.  They are considered separately in \S 6.  The remaining candidates must be screened for indicators of youth.  We follow the procedures outlined in Sh09 and Sh11 and perform a very similar analysis of X-ray and UV flux to identify \emph{PYCs}.

\section{Probable Young Candidates}

\subsection{X-ray Analysis}

We perform a positional cross-correlation of known (T08) and new (LS09, S10) late-type $\beta$ Pic and AB Dor members and our candidate sample with the \emph{ROSAT} All-Sky Survey Bright Source and Faint Source catalogs (\emph{RASS-BSC} and \emph{RASS-FSC}; V99, V00).  All known late-type $\beta$ Pic and AB Dor members and 147 candidates have \emph{ROSAT} counterparts within 50$^{\prime\prime}$.  If a candidate had a counterpart outside of 25$^{\prime\prime}$ they were individually checked for crowded fields and unreliable matches were removed.  X-ray fluxes (\emph{F$_{X}$}) were calculated from count rates and hardness ratios  (Schmitt et al. 1995).  Figure~\ref{xray_plot} shows log(\emph{F$_X$/F$_{K_s}$}) as a function of (\emph{V-K$_{s}$}) color for the previously mentioned subsamples.  We choose to take the ratio \emph{F$_X$/F$_{K_s}$} to be consistent with the use of \emph{K$_{s}$} throughout our analysis, and because \emph{F$_{K_s}$} varies little with activity.  We compare to Sh09, who takes the flux ratio using \emph{F$_J$}, and see an average difference of $\sim$0.1 dex between known \emph{NYMG} member data points.  This reflects the difference in \emph{J} and \emph{K$_s$} band magnitudes, we thus choose log(\emph{F$_X$/F$_{K_S}$}) $\ge$ -2.6 as the cut for \emph{PYCs} to be consistent with Sh09's youth cut and the observed flux ratio difference.  This choice includes $\sim$90$\%$ of the known late-type \emph{NYMG} members.  We identify 114 \emph{PYCs} having \emph{F$_X$/F$_{K_s}$} comparable to or larger than known late-type \emph{NYMG} members.  The remaining 33 candidates that fall below the cut are removed from the sample.   However, the strong X-ray flux detected in some of these \emph{PYCs} may be unreliable as a youth indicator since they are later than mid-M; these stars will be discussed later in this section. 

The dashed grey line in Fig.~\ref{xray_plot} represents a \emph{ROSAT} All-Sky Survey detection limit for the candidate sample.  The detection limit was estimated using the limiting count-rate in the \emph{RASS-BSC} scaled to the source photon limit of the \emph{RASS-FSC}\footnote{The \emph{RASS-BSC} limit is 0.05 cts s$^{-1}$ in the 0.1-2.4 keV energy band, or $\ge$15 source photons in the exposure time.  For the \emph{RASS-FSC} the detection limit is $\ge$6 source photons in the exposure time.}.  This limiting count-rate was combined with an average sample exposure time of $\sim$500 s, the average X-ray hardness ratio (\emph{HR1}) of the known \emph{NYMG} members, and model derived magnitudes from SDF2000 evolutionary models for 40 Myr late-K to mid-M dwarfs at 40 pc to generate the curve.  


\subsection{UV analysis}

The NASA Galaxy Evolution Explorer (\emph{GALEX}) is a space based 0.5m UV telescope sensitive to 1350~\AA~$\le$ $\lambda$ $\le$  2750~\AA~(Martin et al. 2005).  Part of the mission is to produce an All-Sky Imaging Survey (\emph{AIS}) in two bands: the near-UV (\emph{NUV}, 1750 - 2750~\AA) and far-UV (\emph{FUV}, 1350 - 1750~\AA).  Sh11 showed that \emph{GALEX} \emph{NUV} and \emph{FUV} data can be used to effectively identify young M-dwarfs beyond 100 pc, far surpassing the sensitivity of \emph{ROSAT} and providing a new resource in the search for cool, young stars in the solar neighborhood (see also FH10 and R11).  

We used \emph{GalexView}\footnote{a webtool for accessing \emph{GALEX} data available at http://galex.stsci.edu/galexview/} to cross-correlate the known late-type \emph{NYMG} subsample and our candidate list with the sixth data release of the \emph{GALEX} \emph{AIS}, which covers $\sim$75\% of the sky.  We obtained \emph{NUV} and \emph{FUV} data when available for stars with counterparts within 5$^{\prime\prime}$, the resolution of \emph{GALEX} \emph{FUV} channel.  All known \emph{NYMG} members are detected in \emph{NUV} and all except one are detected in \emph{FUV}\footnote{Likely new $\beta$ Pic member PM I04439+3723 is not detected in the \emph{FUV}.  S10 estimate the star to be \emph{SpTy} M3 and lie at a distance of $\sim$80 pc.}.  Three hundred eighty nine candidates have a detection in at least one \emph{GALEX} band.

Fig.~\ref{uv_plot} shows log(\emph{F$_{NUV}$/F$_{K_s}$}) as a function of (\emph{V-K$_{s}$}) for the 304 \emph{NUV} active candidates and known group members.  Comparing to Sh11 we see the same $\sim$0.1 dex difference between our UV/\emph{K$_{s}$} flux ratios and their UV/\emph{J} ratios as in the X-ray analysis.  We thus choose log(\emph{F$_{NUV}$/F$_{K_s}$}) $\ge$ -4.1 and log(\emph{F$_{FUV}$/F$_{K_s}$}) $\ge$ -5.1 as our cuts for \emph{PYCs}, again to be consistent with Sh11 and include the flux difference between \emph{J} and \emph{K$_s$} bands.  We identify 149 \emph{PYCs} with strong \emph{NUV} flux.  Many candidates with strong \emph{NUV} flux also have strong X-ray flux. There are 14 \emph{PYCs} with both \emph{NUV} and \emph{FUV} flux and no \emph{ROSAT} detection that would have been missed if \emph{GALEX} data were unavailable.  Candidates having a \emph{GALEX} \emph{NUV} detection in Fig.~\ref{uv_plot} but falling below the \emph{PYC} cut are removed from the sample.  

The estimated detection limit of the \emph{GALEX} \emph{AIS} is shown as the gray dashed line in Fig.~\ref{uv_plot}.  The limit was estimated using the \emph{GALEX} \emph{AIS}  5$\sigma$ limiting \emph{NUV} magnitude for a 100s exposure (Morrissey et al. 2005) and the same SDF2000 models as the X-ray limit.  Some candidates that we consider \emph{PYCs} due to strong \emph{UV} emission fall below the curve.  This can be attributed to two factors: 1) the exposure time, many of the candidates have actual exposure times $>$100s, 2) the model distance and age; this was chosen for simplicity to be representative of the two \emph{NYMGs}.  Individual candidates have a range of distances and ages.  None the less, the detection limits in both Figs.~\ref{xray_plot} and \ref{uv_plot} correspond to a model derived limiting magnitude of \emph{V}$\sim$16 for the lowest mass candidates.  

\subsection{Undetected Candidates}

Three hundred sixty one candidates are undetected by either \emph{ROSAT} or \emph{GALEX}.~Fig.~\ref{undetected_plot} shows the \emph{V} (top) and \emph{d$_{kin}$} (bottom) distributions of these candidates.  Solid, black and hashed, gray curves represent the undetected candidates (\emph{UCs}) in the \emph{ROSAT} and \emph{GALEX} catalogs respectively.  The \emph{V} distributions show that most \emph{UCs} have 13~$\lesssim$~\emph{V}~$\lesssim$~15 and \emph{d$_{kin}$} $\gtrsim$ 50 pc.  These are most likely main sequence stars with kinematics similar to those of the \emph{NYMGs} selected by chance.  It is likely that the faintest \emph{UCs} are beyond the detection limits of the surveys.   It was expected that $\sim$25\% of the candidate sample would be undetected by \emph{GALEX} because of the sky coverage of the \emph{AIS}.  However, the greater sensitivity of the \emph{GALEX} satellite is apparent from the distributions.  There are fewer undetected candidates overall, the peak of the \emph{V} distribution is shifted to fainter mags by $\sim$1, and the \emph{d$_{kin}$} distribution is approximately flat at large distances.  The faintest \emph{UCs} in the sample could be pursued for individual follow up to check spectroscopic youth indicators, such as gravity sensitive alkali lines, to verify their suspected main-sequence ages.  However, most of the \emph{UCs} appear to be within the estimated limits of the \emph{ROSAT} and \emph{GALEX} surveys and likely represent older stars which only appear to be co-moving with the \emph{BPMG} and \emph{ABDMG} by chance. 

Since the \emph{ROSAT} and \emph{GALEX} catalogs do not have uniform coverage of the entire sky we cannot be certain that we have not missed some potentially young stars but the analyses presented are the best way to identify young stars quickly without intensive observation.  We recognize that the chosen activity indicators may introduce contamination to the \emph{PYC} sample in the form of flaring dwarfs, unresolved white dwarf-M dwarf systems, and unresolved, tidally interacting binaries and that the reliability of these indicators decreases substantially after $\sim$M4 \emph{SpTy}.  To address this problem we devise a simple system to prioritize follow up observations to measure radial velocities and parallaxes.
         

\subsection{Activity, Color, and Priority Indices}

\noindent{\emph{Activity Index, {\tt A}}:  The 204 \emph{PYCs} have varying levels of activity as indicated by their X-ray, \emph{NUV}, and \emph{FUV} flux ratios.  The {\tt A} index shows these variations in a simple way.  It is a three digit binary number where each digit represents whether a youth diagnostic is positive (1) or not (0).  {\tt A}'s first digit represents the X-ray flux ratio, 1 indicates that the ratio is at or above the \emph{PYC} cut, 0 means that the ratio is either below the cut or X-ray flux is not detected.  The second and third digits represent the \emph{NUV} and \emph{FUV} flux ratios respectively.  Their value is selected following the same prescription as the first digit.  Example activity indices for \emph{PYCs} are listed in Table 2.}

\noindent{\emph{Color Index, {\tt C}}:  The reliability of the indicators described by {\tt A} is dependent on \emph{PYC} mass.  W11 used \emph{SDSS} M-dwarf spectroscopic data to show that the fraction of stars with H$\alpha$ emission increases substantially around mid-M type and nearly all late Ms are active.  This is probably the result of convection throughout their interiors and a strengthening of their magnetic dynamos.  Since H$\alpha$ traces activity in a way similar to X-ray and UV emission this activity trend can be applied to the indicators described here.  Thus, {\tt C} is a single digit binary number indicating whether a star has (\emph{V-K$_s$}) $\le$ 5, approximately earlier than M4 according to SDF2000 models, and hence whether {\tt A} is reliable (1) or not (0).  We henceforth designate \emph{PYCs} with {\tt C} = 1 as candidates with reliable youth (\emph{CWRYs}) and candidates with {\tt C} = 0 as candidates with ambiguous youth (\emph{CWAYs}). \emph{PYC} color indices are listed in Table 2.  The \emph{PYC} sample is comprised of 113 \emph{CWRYs} and 91 \emph{CWAYs}.  Even though some \emph{CWAYs} may actually be active older stars they are still worthy of follow up since the activity cuts have been chosen via comparison to known, young, late-type stars.  These candidates are prime targets for studies of gravity sensitive spectral features.}

\noindent{\emph{Priority Index, {\tt P}}:  The observing priority of a \emph{PYC} can be quickly determined by summing the digits of {\tt A} and {\tt C}.  This leads to a new, one digit index {\tt P}, which ranges between 1 - 4 (see Table 2).  The most promising candidates, with strong, reliable indication of youth, have {\tt P} = 4.  We discuss the statistical properties of the \emph{PYCs} and and illustrate the use of their indices when selecting stars for follow up studies in \S \ref{stats}.  

\subsection{Northern PYC List}

The youth indicator analysis has reduced the number of candidates by nearly 70\% to 204 \emph{PYCs}.   Inevitably, it still contains contaminants in the form of active stars whose kinematics align with the \emph{NYMGs} by chance.  Their removal will require follow up observations that are beyond the scope of this paper.  Three stars are candidates for both the \emph{BPMG} and \emph{ABDMG}.  Dual candidacy was expected for some stars due to the overlap of the two groups in \emph{U} and \emph{W} velocity space. We present an example of the format of the northern \emph{PYC} list in Table 2 and explain the column designations in the following subsections.  The entire list is available in the electronic version of the journal in machine readable format.

Columns 1, 2, and 3 list our internal ID and \emph{Hipparcos} and \emph{Tycho-2} IDs if available.  The internal ID is based on the star's coordinates in the \emph{ICRS} epoch 2000.0 system.  The final letter (N, S, E, W) is optional and distinguishes pairs of stars that would otherwise have the same coordinates. The choice depends on the orientation of the pair.  If the separation in declination is larger, the stars are appended N and S.  If the separation in right ascension is larger, the pair is given E and W.  Candidates with NS or EW are not necessarily common proper motion pairs, although they may be.  If only one star appears with a directional designation, its counterpart was cut from the sample during the search, usually because it fails to meet the (\emph{V-K$_s$}) cut. 

Columns 4 and 5 list the coordinates in the \emph{ICRS} epoch 2000.0 system.  Columns 6 - 9 give the \emph{V}, \emph{J}, \emph{H} and \emph{K$_{s}$} photometry of the candidate.  If the star is in the \emph{Hipparcos} or \emph{Tycho-2} catalog the \emph{V} mags from those sources are used. If the star is missing from those catalogs the \emph{V} mag is a combination of the USNO-B1.0 (Monet et al. 2003) \emph{B$_{J}$},  \emph{R$_{F}$}, and \emph{I$_{F}$} photographic magnitudes following the conventions of LS05.  \emph{J}, \emph{H} and \emph{K$_{s}$} magnitudes are from \emph{2MASS}.  Columns 10 and 11 list the proper motions of the \emph{PYCs}.  Proper motions are taken from the \emph{Hipparcos} and \emph{Tycho-2} catalogs if the candidate has a counterpart, otherwise the \emph{LSPM/SBK} derived proper motions are used.  We adopt the $\pm$8 mas yr$^{-1}$ uncertainty of the \emph{LSPM} and \emph{SBK} catalogs as the proper motion uncertainty in the list.
   
Columns 12 and 13 list the \emph{d$_{kin}$} of the candidate and its uncertainty.  Column 14 and 15 list the predicted radial velocity (\emph{RV$_{p}$}, see LS09 and S10) of the candidate and its uncertainty which includes the intrinsic spread in \emph{BPMG} or \emph{ABDMG} \emph{UVW} velocities quoted in \S 2.   Columns 16, 17, and 18 list the logarithm of the X-ray, \emph{NUV}, and \emph{FUV} flux ratios described in \S 4.  Columns 19, 20, and 21 are the activity ({\tt A}), color ({\tt C}) and priority ({\tt P}) indices respectively (see \S 4) and column 22 identifies whether a star is a \emph{PYC} of the \emph{BPMG} (1) or the \emph{ABDMG} (2).  Stars with dual candidacies are listed with identical IDs, coordinates, proper motions, and photometry and differ only in their \emph{d$_{kin}$} and \emph{RV$_{p}$}, which are specific to the particular \emph{NYMG}.

\section{Statistical Properties}\label{stats}

The \emph{PYC} sample is comprised of 49 \emph{BPMG} and 155 \emph{ABDMG} candidates.  Figure~\ref{PYChist_plot} shows the distributions of the \emph{PYC} \emph{V} mags and \emph{d$_{kin}$} with the \emph{BPMG} and \emph{ABDMG} \emph{PYCs} shown in hashed, gray and solid, black histograms respectively.   In the \emph{V} mag distributions \emph{BPMG} candidates appear brighter on average.  This is most likely due to younger age of the group.  The \emph{BPMG} distance distribution also exhibits many fewer candidates at large \emph{d$_{kin}$}.  Furthermore, the distribution is flat after $\sim$50 pc.  Brighter candidates at smaller estimated distances make \emph{BPMG} \emph{PYCs} generally more favorable for follow up observations.

\subsection{Index Distributions}

In this subsection we break the \emph{PYC} sample into subgroups as defined by the {\tt A}, {\tt C}, and {\tt P} indices.  We first discuss the color index ({\tt C}) subgroups and compare their cumulative activity index ({$\Sigma$\tt A}) distributions.  Fig.~\ref{CWRY_plot} shows the \emph{V} mag (top) and \emph{d$_{kin}$} (bottom) distributions for \emph{CWRYs}.  The three histograms plotted represent \emph{CWRYs} with $\Sigma${\tt A} = 1 (solid, gray), 2 (hashed, dark gray), or 3 (open, red).  The largest contribution to the \emph{CWRY} sample are candidates having only one strong activity indicator.  These candidates typically have only strong X-ray or \emph{NUV}.  It is striking that there are more \emph{CWRYs} with $\Sigma${\tt A}=3, those with the strongest indication of youthful activity, than $\Sigma${\tt A}=2.  The $\Sigma${\tt A} = 3 distance distribution also peaks at smaller \emph{d$_{kin}$}.  The larger number of $\Sigma${\tt A}=3 candidates at smaller predicted distances makes follow up of these \emph{CWRYs} particularly pertinent.

The same distributions using the same color designations in Fig.~\ref{CWRY_plot} for the \emph{CWAYs} are shown in Fig.~\ref{CWAY_plot}.  The \emph{V} distributions are shifted $\sim$2 mags fainter than the \emph{CWRYs}.  This is expected since \emph{CWAYs} are systematically redder and represent lower-mass candidates.   Like in the \emph{CWRY} distributions, candidates with $\Sigma${\tt A} = 1 dominate, but to a larger degree.  However, the number of \emph{CWRYs} with $\Sigma${\tt A} = 2 and $\Sigma${\tt A} = 3 is nearly equal.  The \emph{d$_{kin}$} distribution of $\Sigma${\tt A} = 3 \emph{CWAY} sample is approximately flat.  This feature is inconsistent with the other distance distributions in the figure, which have significant peaks at \emph{d$_{kin}$} $\ge$ 60 pc.  The flat distance distribution is indication that contamination does not increase as volume increases in the $\Sigma${\tt A} = 3 \emph{CWAY} sample.  Thus these candidates are also prime for follow up.  

As a final guide to prioritize follow up, Fig.~\ref{P_plot} shows the {\tt P} index \emph{V} mag and \emph{d$_{kin}$} distributions of the young candidates.  The {\tt P} = 1 sample consists entirely of candidates with (\emph{V-K$_s$}) $>$ 5.0 (\emph{CWAYs}) that have only one strong activity indicator.  These candidates are the lowest priority for immediate \emph{RV} follow up, but are prime targets for the study of gravity sensitive spectral features.  The {\tt P} = 2 and 3 distributions are a mixture of \emph{CWRYs} and \emph{CWAYs} that may have multiple strong activity indicators.  The {\tt P} = 4 distribution consists only of candidates with (\emph{V-K$_s$}) $\le$ 5.0 (\emph{CWRYs}) that have strong X-ray, \emph{NUV}, and \emph{FUV} flux.  These candidates have the strongest indication of reliable youth and are the highest priority for follow up.  The {\tt P} = 3 and 4 distance distributions (Fig.~\ref{P_plot}, bottom) do not have strong peaks beyond 60 pc. This indicates a lower rate of contamination in these samples as volume increases.
 
\subsection{Survey Completeness and Moving Group Densities}

Our search for low-mass stars in the \emph{BPMG} and \emph{ABDMG} is effectively
complete for stars within 70 pc to within the proper motion limit of
the \emph{SBK} catalog. Because the \emph{SBK} catalog is complete to
V $\approx$ 20 and has a proper motion limit $\mu>$~40 mas yr$^{-1}$, all
low-mass stars within 70 pc will be detected if their absolute V magnitude \emph{$M_V$} $<$ 15.5 and if their velocity in the plane of the sky
$v_{tran}$ $>$ 13.3 km s$^{-1}$. Regardless of how much contamination
there is in our sample, the fact remains that virtually all moving
group members that have absolute magnitudes and velocities within the
limits above are to be found in our list.

Not all stars in the \emph{BPMG} and \emph{ABDMG} will have proper motions large
enough to be detected. To estimate the fraction of stars detected as a
function of their distance, we have performed a simple Monte-Carlo
simulation which assumes a uniform density of moving group members in
the solar vicinity, and estimates how many stars in the synthetic
sample would have proper motions large enough to be detected. In
addition to the kinematic selection, we add a 5\% probability that any
star may be overlooked, consistent with the estimated $>95$\%
completeness of the \emph{SBK} catalog. Results are displayed in Fig.~\ref{comp_plot}
(upper panels) for both the \emph{BPMG} and \emph{ABDMG}; one finds that $>$75\%
of the \emph{BPMG} stars at $\approx$70~pc from the Sun should be on our list,
and larger fractions of more nearby objects for a total of
$\approx$86.8\% of all \emph{BPMG} within 70~pc. For the \emph{ABDMG}, the search is
significantly more complete due to the larger systemic velocity of the
group, which yields larger proper motions on average than \emph{BPMG}
stars. The simulation suggests that $>$85\% of the \emph{ABDMG} low-mass
stars at $\approx$70~pc should be in the list, and $\approx$94.4\% of all
the members within that range.

With this high level of completeness, it is possible to place firm
upper limits on the volume density of \emph{BPMG} and \emph{ABDMG} stars in the
northern hemisphere based on the number of stars we identified. We
separate the candidates in 10~pc distance bins, count the number of
stars with absolute magnitudes 8.0$~<M_V<~$15.5 within each bin, and
divide this number by the detected fraction as estimated above. We
then calculate the density of stars $\rho_*$ in that distance range by
dividing by the volume enclosed by each of the corresponding
hemispheric shell, and express the result in stars per cubic parsec
(stars pc$^{-3}$). Results are displayed in Fig.~\ref{comp_plot} (lower
panels). We calculate $\rho_*$ as a function of distance for the
complete list of low-mass star candidates (open triangles) and for
the subset of candidates which show strong evidence of youth (\emph{PYCs}, filled
squares).

If we believe that all group members should display indicators of
youth, while allowing that some non-members might show signs of youth
as well, then the mean density of young candidates sets an upper limit
for the density of the group. Under this assumption, we find that in
the northern sky volume within 70~pc of the Sun, the \emph{BPMG} has a maximum
number density of low-mass stars of $\approx$6.7~x~10$^{-5}$ stars
pc$^{-3}$, which means that the same volume should be host to at most
$\approx$36 low-mass (M dwarf) members of the \emph{BPMG}. For the \emph{ABDMG}, we
find that the group has a maximum density of $\approx$2.2~x~10$^{-4}$
stars pc$^{-3}$ for low-mass stars, which suggests that the 70~pc
volume should contain at most $\approx$119 low-mass (M dwarf)
members.  We expect that most of the true low-mass, north-sky members of the \emph{BPMG}
and \emph{ABDMG} should be in our Table 2.

\section{Preliminary Results} 

\subsection{Ultracool Candidates}

Seven candidates in Fig.~\ref{ccd_plot} are redder in (\emph{H-K$_{s}$}) than the expected dwarf sequence (black box).  These stars are potential ultracool candidates with \emph{SpTy's} M7 and later.  We perform a literature search for pertinent youth and kinematic data regarding these candidates since they are not expected to exhibit strong activity.  The literature reveals three \emph{LNMs}, another three stars that require further follow up, and one that is ruled out on the basis of inconsistent \emph{RVs} (see Table~3).       

\subsection{\emph{PYCs} with \emph{Hipparcos} Distances}

Five stars in the \emph{PYC} sample have distances measured by \emph{Hipparcos}, this allows us to directly test group membership by verifying that  \emph{d$_{kin}$} is consistent with the parallax distance (\emph{d$_{\pi}$}).  We have calculated distances for all stars listed in the \emph{Hipparcos} catalog based on measured parallaxes from the recent new reduction by van Leeuwen (2007).  Table~4 compiles literature data on these 5 stars and lists their ID, \emph{Hipparcos} ID, coordinates in the \emph{ICRS} system, \emph{V} mags, \emph{2MASS} K$_{s}$ mags, predicted \emph{d$_{kin}$} and measured \emph{d$_{\pi}$} distances, predicted and measured \emph{RV}, \emph{NYMG} candidacy, and if the star was removed from the sample.  5 \emph{LNMs} (1 triple system and 2 singles) are identified, and 2 are ruled out on the basis of inconsistent measured \emph{RV} (see Table~4).  All of the candidates with \emph{Hipparcos} distances are \emph{CWRYs} with {\tt P} = 3 or 4.

\subsection{Likely New Members}

\emph{PYC J09362+3731 = HIP 47133}:  HIP 47133 is a late-K dwarf at a distance of 31 pc, consistent with our predicted distance.  The star exhibits strong \emph{NUV}, and \emph{FUV} emission, consistent with known late type \emph{BPMG} members.  We identify HIP 47133 as a likely new member of the \emph{BPMG} on the basis of consistent distance and strong activity.  \emph{RV} measurements should be performed to confirm membership.

\emph{PYC J10143+2104 = HIP 50156}:  HIP 50156 is an M1 dwarf at a distance of $\sim$20 pc.  
Strong X-ray flux, moderate H$\alpha$ emission, possible low surface gravity from CaH indices, but no 6708~\AA~lithium line are described by 
Sh09.   We identify strong \emph{NUV} and \emph{FUV} flux as well.  
L\'opez-Santiago et al. (2010, hereafter L10) measure an \emph{RV} of 2.7$\pm$1.0 km s$^{-1}$, which matches well 
with our predicted \emph{RV}, and calculate (\emph{UVW}) = (-6.8, -18.3, -8.2).  L10 classify the star as a member of the local association and also report a Li non-detection.  However, the (\emph{UVW}) velocities are more consistent with those of the \emph{BPMG} than those of the local association (Montes et al. 2001).  The lack of detected Li in both Sh09 and L10 is contrary to expectations for $\sim$10 Myr old early-M dwarfs.  But, BC10 suggest that Li content is not a reliable youth indicator since it is strongly affected by stellar accretion history.  We therefore identify HIP 50156 as a likely member of the \emph{BPMG} based on consistent \emph{UVW} kinematics and strong activity and suggest further follow up for confirmation.

\emph{PYC J01036+4051 = HIP 4967}:  HIP 4967 is a visual triple system at a distance of 
$\sim$30 pc.  Sh09 identify indicators of youth in the spectra of the M3 
secondary and M4 tertiary components, including H$\alpha$ emission, and estimate the ages 
to be 20 - 300 Myr.  We also identify strong activity in components of the system.
Based on evidence of youth and consistent distance we identify HIP 4967 as a likely member 
of the \emph{ABDMG}.  \emph{RV} measurements should be pursued to verify membership.  

\emph{PM  I00194+4614 = 2MASS J0019262+461407}:  \emph{2MASS} J0019262+461407 (hereafter 2M0019) was identified by Cruz et al. (2003) as an M8 dwarf in a search for nearby ultracool objects in the \emph{2MASS} catalog.  Reiners \& Basri (2009, 2010, hereafter RB09,10) measure an \emph{RV} of -19.5$\pm$2.0 km s$^{-1}$, which is consistent with our predicted \emph{RV}, and a remarkably large v$sin$i of 68$\pm$10 km s$^{-1}$.  In their high resolution optical spectra they also identify lithium at 6708~\AA~in absorption, indicating that 2M0019 is most likely a young brown dwarf (M $\lesssim$ 65 M$_{Jup}$, Age $<$ 0.5 Gyr).  We identify 2M0019 as a \emph{LNM} of the \emph{ABDMG} based on indicators of youth and consistent \emph{RV}.  Assignment as a true group member awaits parallax measurement to verify that d$_{kin}$ is the true distance and a more detailed analysis of youth indicators.

\emph{PM I04436+0002 = 2MASS J0443376+000205}:  \emph{2MASS} J0443376+000205 (hereafter 2M0443) was identified as an M9 dwarf by Hawley et al. (2002).  Cruz et al. (2007) identified indicators of low surface gravity in its spectrum and propose that the object is a young brown dwarf.  The \emph{RV} and v$sin$i of 2M0443 were measured by RB09 and RB10 to be 17.1$\pm$2.0 km s$^{-1}$ and 13.5$\pm$2.0 km s$^{-1}$ respectively.  The \emph{RV} measurement is consistent with our predicted \emph{RV}.  Reiners and Basri also identify lithium absorption in the dwarf, strengthening the argument that 2M0443 is a young brown dwarf.  Thus, we identify 2M0443 as a \emph{LNM} of the \emph{ABDMG} and will pursue a parallax measurement to verify membership.  

\emph{PM I13143+1320 = 2MASS J1314203+132001}:  \emph{2MASS} J1314203+132001 (hereafter 2M1314) was first identified as the high proper motion star NLTT 33370 by Luyten (1979).  Based on its photometry, LS05 estimate its distance to be 9.7$\pm$3.0 pc.  L\'epine et al. (2009, hereafter L09) measure strong H$\alpha$ emission (EW = 54.1~\AA) and a parallactic distance of 16.4$\pm$0.8 pc. This is nearly twice the photometric distance but in agreement with our predicted \emph{d$_{kin}$}.  Since \emph{d$_{phot}$} significantly underestimates the true distance to 2M1314 the star could be an unresolved binary or very young.  Either of these scenarios  can be consistent with the strong H$\alpha$ emission.  If 2M1314 is $\lesssim$100 Myr old its mass would be very close to the substellar limit.  We identify 2M1314 as a \emph{LNM} of the \emph{ABDMG} based on the evidence in the literature and will pursue \emph{RV} measurements and follow up on further indicators of youth in order to verify group membership.          

If these three ultracool \emph{LNMs} are verified to be bona fide members of the AB Dor group they will be the lowest mass, isolated members and represent important benchmarks.

\section{Summary}

We have presented a list of 204 young moving group candidates in the northern hemisphere.  These stars are candidates of the $\sim$10 Myr old $\beta$ Pictoris moving group and $\sim$70 Myr old AB Doradus moving group.  Candidates were selected from the \emph{Tycho-2}, \emph{LSPM-North}, and \emph{SBK} catalogs using the search algorithm first described in LS09 and S10.  Selection criteria used in choosing candidates were derived from the known members of the \emph{NYMGs}.  X-ray, \emph{NUV}, and \emph{FUV} activity indicators were used to trim the initial sample to only the young stars following the convention of Sh09 and Sh11.  Identification of true moving group members in the list requires \emph{RV} and parallax measurements to verify consistent group kinematics.  We have devised a simple system of indices based on activity and color to help prioritize follow up observations.  

From detailed inspection and data gathered in the literature, we can identify 2 \emph{LNMs} of the \emph{BPMG} and six of the \emph{ABDMG} (3 in a triple system).  Three of the \emph{LNMs} of the \emph{ABDMG} are possible young brown dwarfs.  The identification of 3 \emph{LNMs} as possible brown dwarfs is particularly valuable because if follow up proves them true members of the group they will be the lowest-mass, isolated members of the group yet identified.  Not only are these objects interesting for assessing the substellar mass function of the \emph{ABDMG}, they may represent benchmark young substellar objects which are crucial for understanding substellar formation and evolution.  Our proper motion search technique has now yielded at least 24 likely new moving group members (LS09, S10, this work).

This paper concerns only candidates of the \emph{BPMG} and \emph{ABDMG} north of the celestial equator.  Since 80 to 90\% of the known $\beta$ Pic and AB Dor group members lie in the south (see T08), we can expect that extension of our identification procedures to the south will yield similarly greater numbers of \emph{PYCs} and \emph{LNMs}.  A complete census of \emph{NYMG} members including the low-mass stars will allow for the study of: 1) the origins of the moving groups, 2) stellar evolution in a relatively nearby sample of \emph{PTTSs} and young brown dwarfs and, 3) exoplanets and their formation. 

\
\
\
\
\acknowledgments

We thank the referee for helpful comments that improved this manuscript.  We thank C. Bender for sharing with us his software for the extraction of CSHELL spectra and for their analysis by correlation techniques.  The work of J.E.S and M.S. was supported in part by NSF grant AST 09-07745.  The work of S.L. was supported by NSF grants AST 06-07757 and AST 09-08406. This publication makes use of data products from the Two Micron All Sky Survey, which is a joint project of the University of Massachusetts and the Infrared Processing and Analysis Center/California Institute of Technology, funded by the National Aeronautics and Space Administration and the National Science Foundation.  This research has made use of the SIMBAD database, Aladin, and Vizier, operated at CDS, Strasbourg, France.

\clearpage

\begin{sidewaystable}
\hspace{-1.5cm}
\begin{tabular}{lllrrccccr}
\multicolumn{10}{c}{\textbf{Table 1}}\\
\multicolumn{10}{c}{\textbf{Recovered Known NYMG Members}}\\
\hline
\hline
Cat ID			&	\emph{Hip.} ID	&	\emph{Tycho2} ID 	&$\alpha(\emph{ICRS})$	& $\ $ $\delta(\emph{ICRS})$	& \emph{V}            	&\emph{K$_{s}$}  	& \emph{d$_{kin}$} 	& \emph{SpTy}$^{a}$	& Membership\\
		              	&			     	&				&(2000.0)$\ $    	&  (2000.0)$\ $            	& (mag) 		& (mag)   		& (pc)                     	&           		&                        \\   
\hline
PM I02414+0559	&12545			&                            	& 40.357864            	& 5.988454                  	&10.3       		& 7.1            	& 38.5$\pm$3.7    	& K6Ve		&\emph{BPMG}\\	
PM I04595+0147	&23200	     		&85 1075 1	         &74.895126	         &1.783535	           	&10.3	 	&6.3 	    		&25.8$\pm$2.9		& M0Ve		&\emph{BPMG}\\
PM I17386+6114	&86346	     		&4199 1286 1		&264.665100	         &61.237804	          	&10.3	 	&6.8             	&24.2$\pm$5.1		&K7Ve		&\emph{ABDMG}\\
PM I21521+0537	&107948	     		&556 1289 1		&328.043396	         &5.626641	          	&12.1	 	&7.4	             	&29.9$\pm$2.0		&M2Ve		&\emph{ABDMG}\\
PM I22234+3227	&110526	     		&2738 1390 1		&335.871185	         &32.459461	          	&10.7	 	&6.1	             	&15.0$\pm$1.1		&M3e		&\emph{ABDMG}\\
PM I23060+6355	&114066	     		&4286  212 1		&346.520111	         &63.926220	          	&10.9	 	&7.0	             	&23.7$\pm$2.0		&M1e		&\emph{ABDMG}\\
\hline
\multicolumn{10}{l}{\scriptsize{$^{a}${\emph{SpTy's} from T08}}}\\
\label{known_mem}
\end{tabular}
\end{sidewaystable}

\clearpage

\begin{sidewaystable}
\tiny
\hspace{-2.6cm}
\begin{tabular}{lllrrccccccccccccccccc}
\multicolumn{22}{c}{\textbf{Table 2}}\\
\multicolumn{22}{c}{\textbf{Northern PYC List}}\\
\hline
\hline
PYC ID			&\emph{Hip.} ID		&\emph{Tycho2} ID		&$\alpha(\emph{ICRS})$	&$\ $ $\delta(\emph{ICRS})$			& \emph{V}  	& \emph{J} 	& \emph{H}           	&\emph{$K_{s}$}  & $\mu$$_{\alpha}$	&$\mu$$_{\delta}$  	& \emph{d$_{kin}$}	& $\delta$\emph{d$_{kin}$}	&\emph{RV$_{p}$}	&$\delta$\emph{RV$_{p}$}  &$\ $[\emph{X}]$^{a}$    &$\ $[\emph{N}]$^{b}$ 	&$\ $[\emph{F}]$^{c}$        & {\tt A}  &{\tt C}  &  {\tt P} &  Cand.$^d$\\ 
			&			&				& (2000.0)  		&(2000.0)					&(mag)	&(mag)	&(mag)		&(mag)	  &(mas yr$^{-1}$)		         &(mas yr$^{-1}$)      	         &(pc)			&(pc)			         &(km s$^{-1}$)       	&(km s$^{-1}$)	                    &(dex)                                &(dex)                                            &(dex)                                        &           &          &   &\\
\hline
J00025+3422N &             &                             &0.645568   &34.381161  &13.6  &10.4   &9.7   &9.5     &67.0    &-52.0  &68.8   &7.4  &-14.11   &1.31   &0.00  &-3.80  &-4.23  &011 &1 &3  &2\\
J00197+1951  &               &                             &4.929336   &19.853241  &15.7  &10.7  &10.1   &9.9     &59.0    &-45.0  &59.4   &7.9   &-1.73   &1.05  &-1.74  &-3.86   &0.00  &110 &0 &2  &1\\
J00270+6630  &               &                             &6.761791   &66.510834  &13.2   &9.6   &8.9   &8.7     &66.0    &-23.0  &56.9   &8.3   &-9.16   &1.27  &-2.36   &0.00   &0.00  &100 &1 &2  &1\\
J00270+6630  &               &                             &6.761791   &66.510834  &13.2   &9.6   &8.9   &8.7     &66.0    &-23.0  &69.5   &9.0  &-20.81   &1.23  &-2.36   &0.00   &0.00  &100 &1 &2  &2\\
J00325+0729  &               &                             &8.145008    &7.490874  &12.8   &8.4   &7.8   &7.5     &92.0    &-55.0  &41.1   &4.4    &1.70   &1.05  &-1.60  &-3.86  &-4.47  &111 &0 &3  &1\\
J00390+1330  &               &                             &9.764313   &13.504687  &15.7  &10.9  &10.4  &10.1    &66.0    &-70.0  &67.7   &6.4   &-3.45   &1.45  &-1.97  &-3.79  &-3.96  &111 &0 &3  &2\\
J00411+5523  &               &                             &10.296559   &55.399475  &16.6  &11.4  &10.8  &10.6     &84.0    &-61.0  &51.7   &4.9  &-17.76   &1.24  &-1.60   &0.00   &0.00  &100 &0 &1  &2\\
J00450+2634  &               &                             &11.250682   &26.569508  &12.5   &9.5   &8.9   &8.7     &85.0   &-100.0  &48.2   &3.7   &-8.06   &1.37  &-2.02  &-3.65   &0.00  &110 &1 &3  &2\\
J00484+3632  &               &2288   758  1     &12.119358   &36.542927  &11.7   &9.2   &8.6   &8.5     &59.0    &-40.0  &61.4   &8.4   &-3.14   &1.25   &0.00  &-3.90   &0.00  &010 &1 &2  &1\\
J00489+4435  &               &                             &12.242649   &44.585823  &13.9   &9.1   &8.5   &8.2    &116.0   &-136.0  &32.6   &2.2  &-14.15   &1.27  &-1.65   &0.00   &0.00  &100 &0 &1  &2\\
J00495+0356  &               &                             &12.394575    &3.938275  &14.8  &10.9  &10.3  &10.0     &71.0    &-88.0  &57.9   &4.8    &1.35   &1.51   &0.00  &-3.90   &0.00  &010 &1 &2  &2\\
J00546+3434  &               &                             &13.671499   &34.575741  &15.4  &10.9  &10.3  &10.1     &60.0    &-68.0  &68.1   &6.8  &-10.30   &1.32   &0.00  &-3.89   &0.00  &010 &0 &1  &2\\
J01028+1856  &               &                             &15.712469   &18.948402  &14.4   &9.5   &8.9   &8.7     &94.0    &-53.0  &40.9   &4.4    &1.53   &1.23   &0.00  &-3.90  &-4.36  &011 &0 &2  &1\\
J01036+4051   &4967      &2803   800  1   &15.917160   &40.858143  &10.9   &8.1   &7.5   &7.3    &122.0   &-167.0  &29.2   &1.8  &-11.93   &1.29  &-2.05  &-3.51  &-4.38  &111 &1 &4  &2\\
J01037+4051W &            &                            &15.925412   &40.854427  &13.5   &9.4   &8.8   &8.5    &132.0   &-164.0  &28.7   &1.8  &-11.93   &1.29  &-1.57  &-3.83  &-4.44  &111 &1 &4  &2\\
\hline
\multicolumn{22}{l}{\tiny{Note:  This table is available in its entirety in machine readable format in the online version of the journal.  This portion is given as an example of format and content.}}\\
\multicolumn{22}{l}{\tiny{$^a$log(\emph{F$_X$/F$_{Ks}$}).}}\\
\multicolumn{22}{l}{\tiny{$^b$log(\emph{F$_{NUV}$/F$_{Ks}$}).}}\\
\multicolumn{22}{l}{\tiny{$^c$log(\emph{F$_{FUV}$/F$_{Ks}$}).}}\\
\multicolumn{22}{l}{\tiny{$^d$Candidate of (1) \emph{BPMG}, (2) \emph{ABDMG}.}}\\
\label{PYC_list}
\end{tabular}
\end{sidewaystable}

\begin{sidewaystable}
\tiny
\hspace{-2.6cm}
\begin{tabular}{llrrcccccccccccl}
\multicolumn{16}{c}{\textbf{Table 3}}\\
\multicolumn{16}{c}{\textbf{Ultracool Candidates}}\\
\hline
\hline
Cat ID        	&	\emph{2MASS} ID 		&$\alpha$(\emph{ICRS})$\ $	&$\delta$(\emph{ICRS})$\ $	& \emph{V}  	& \emph{J} 	& \emph{H}           &\emph{K$_{s}$}  	& \emph{d$_{kin}$}	& \emph{d$_{\pi}$}		& \emph{d$_{phot}$}		& \emph{RV$_{p}$} 	& \emph{RV$_{m}$}		& \emph{SpTy}		& Cand. 	& Notes\\  				
		       	&			  		&(2000.0)$\ $            &  (2000.0)$\ $    		& (mag)     	& (mag) 	& (mag)               & (mag)          	& (pc)        	& (pc)        		& (pc)        		& (km s$^{-1}$) 	& (km s$^{-1}$)    	 	&                  	&       		&   \\   					        
\hline
\multicolumn{16}{c}{\textbf{Likely New Members}}\\
\hline
PM I00194+4614	&J0019262+461407		&4.859472		&46.235477			&20.0	&12.6	&11.9		&11.5		&39.6$\pm$3.1		&					&19.5$\pm$1.6$^{a}$	&-16.5$\pm$1.3	&-19.5$\pm$2.0$^{b}$	&M8$^{a}$	&\emph{ABDMG} 		&RV$_{p}$~$\approx$~RV$_{m}$, youth$^{b,c}$\\
PM I04436+0002	&J0443376+000205		&70.906723		&0.034731			&19.6	&12.5	&11.8		&11.2		&39.5$\pm$3.1		&					&16.2$\pm$1.0$^{d}$	&19.0$\pm$1.4		&17.1$\pm$2.0$^{b}$	&M9$^{d}$	&\emph{ABDMG} 		&RV$_{p}$~$\approx$~RV$_{m}$, youth$^{b,d}$  \\
PM I13143+1320	&J1314203+132001		&198.584839		&13.333536			&15.9	&9.8		&9.2			&8.8			&20.1$\pm$1.0		&16.4$\pm$0.8$^{g}$	  &9.7$\pm$3.0$^{g}$  	&-10.3$\pm$1.6	&					&M7$^{g}$	&\emph{ABDMG} 		&d$_{kin}$~$\approx$~d$_{\pi}$, youth$^{g}$\\	
\hline
\multicolumn{16}{c}{\textbf{Require Further Followup}}\\
\hline
PM I03010+4416	&J0301032+441656		&45.263393		&44.282410			&18.0	&12.1	&11.4		&11.0		&50.6$\pm$3.9		&					&24.5$\pm$4.2$^{a}$	& -6.4$\pm$1.3		&					&M6$^{a}$	&\emph{ABDMG}		&\\					
PM I09069+0301	&J0906955+030117		&136.733276		&3.021399			&14.6	&11.1	&10.5		&10.1		&65.6$\pm$6.8		&					&					&14.0$\pm$1.4		&					&			&\emph{ABDMG}		&\\           
PM I15291+6312	&J1529101+631253		&232.292389		&63.214989			&19.5	&11.6	&10.9		&10.6		&21.7$\pm$2.4		&					&					&-27.6$\pm$1.4	&					&			&\emph{ABDMG}		& \\			
\hline
\multicolumn{16}{c}{\textbf{Ruled Out}}\\
\hline
PM I08109+1420	&J0810586+142039		&122.744370		&14.344112			&19.6	&12.7	&12.1		&11.6		&46.0$\pm$3.6		&					&13.8$\pm$2.1$^{e}$	&11.9$\pm$1.3		&27.4$\pm$0.4$^{e}$	&M9$^{f}$		&\emph{ABDMG} 		&RV$_{p}$~$\not\approx$~RV$_{m}$\\					

\hline
\multicolumn{16}{l}{\tiny{$^{a}${Cruz et al. (2003)}}}\\
\multicolumn{16}{l}{\tiny{$^{b}${A. Reiners \& G. Basri (2009)}}}\\
\multicolumn{16}{l}{\tiny{$^{c}${A. Reiners \& G. Basri (2010)}}}\\
\multicolumn{16}{l}{\tiny{$^{d}${Cruz et al. (2007)}}}\\
\multicolumn{16}{l}{\tiny{$^{e}${Reid et al. (2002)}}}\\
\multicolumn{16}{l}{\tiny{$^{f}${Gizis et al. (2000)}}}\\
\multicolumn{16}{l}{\tiny{$^{g}${L\'epine et al. (2009)}}}\\
\label{ucool_cands}
\end{tabular}
\end{sidewaystable}

\begin{sidewaystable}
\scriptsize
\hspace{-2.5cm}
\begin{tabular}{llrrcccccccll}
\multicolumn{13}{c}{\textbf{Table 4}}\\
\multicolumn{13}{c}{\textbf{Candidates With Consistent Hipparcos Distances}}\\
\hline
\hline
PYC ID			&\emph{Hip.} ID	&$\alpha$(\emph{ICRS})	&$\delta$(\emph{ICRS})$\ $	& \emph{V}	&\emph{$K_{s}$}	& \emph{d$_{kin}$}	&\emph{d$_{\pi}$}$^{a}$	& \emph{RV$_{p}$}	& \emph{RV$_{m}$}		& \emph{SpTy}			&Cand.	&Notes\\
				&			&(2000.0)$\ $        	&  (2000.0)$\ $            	& (mag)    	& (mag)          	& (pc)            	& (pc)        		& (km s$^{-1}$) 	& (km s$^{-1}$)    	 	&                  		&    			&\\   
\hline   
\multicolumn{13}{c}{\textbf{Likely New Members}}\\
\hline
J01036+4051	&4967		&15.917160		&40.858143			&10.9	&7.3			&29.2$\pm$1.8		&29.9$\pm$2.1			&-11.9$\pm$1.3	&					& $\sim$M$^{b}$   	&\emph{ABDMG}		& VT$^{b}$\\ 
J09362+3731	&47133		&144.066330		&37.529320			&11.1	&7.2			&33.0$\pm$2.9		&33.7$\pm$2.6			&1.0$\pm$1.6		&					&K5$^{c}$			&\emph{BPMG}		&\\
J10143+2104	&50156		&153.579926		&21.074898			&10.0	&6.3			&20.7$\pm$1.7		&23.1$\pm$1.0			&2.8$\pm$1.3		&2.7$\pm$0.1$^{d}$		&M1$^{b}$		&\emph{BPMG}		&RV$_{p}$~$\approx$~RV$_{m}$\\
\hline
\multicolumn{13}{c}{\textbf{Ruled Out}}\\
\hline
J12198+5246	&60121		&184.950302		&52.779182			&11.1	&7.5			&25.0$\pm$1.6		&28.0$\pm$1.7			&-18.6$\pm$1.5	&-4.9$\pm$1.2$^{e}$	&K7$^{c}$			&\emph{ABDMG}		&RV$_{p}$~$\not\approx$~RV$_{m}$\\
J12576+3513S	&63253		&194.417801		&35.225014			&10.6	&6.6			&17.6$\pm$0.9		&19.3$\pm$1.1			&-16.3$\pm$1.6	&-9.5$\pm$0.6$^{f}$	&M4$^{b}$		&\emph{ABDMG}		&RV$_{p}$~$\not\approx$~RV$_{m}$, VB$^{b}$\\ 
\hline
\multicolumn{13}{l}{\scriptsize{$^{a}$distance from \emph{Hipparcos} new reduction parallax (van Leeuwen 2007)}}\\
\multicolumn{13}{l}{\scriptsize{$^{b}$VT = visual triple, VB = visual binary; Shkolnik et al. (2009)}}\\
\multicolumn{13}{l}{\scriptsize{$^{c}$SpTy from SIMBAD}}\\
\multicolumn{13}{l}{\scriptsize{$^{d}$L\'opez-Santiago et al. (2010)}}\\
\multicolumn{13}{l}{\scriptsize{$^{e}$Upgren et al. (1996)}}\\
\multicolumn{13}{l}{\scriptsize{$^{f}$This work, using CSHELL at the NASA-IRTF (Greene et al. 1993)}}\\
\label{hip2_table}
\end{tabular}
\end{sidewaystable}

\begin{figure}
\epsscale{1.0}
\plotone{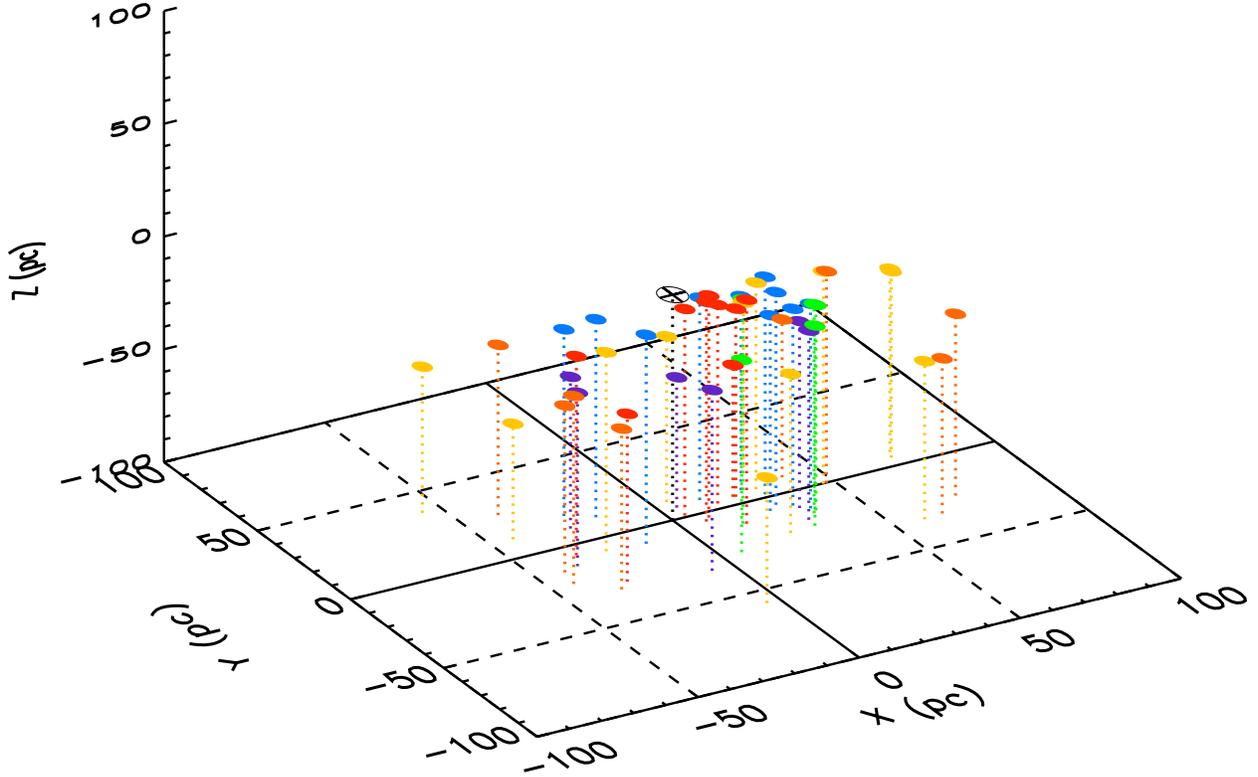}
\caption{\emph{BPMG} \emph{XYZ} distribution, colors represent increments of (\emph{V-K$_s$}) color, a rough indicator of mass.  From Siess et al. (2000) models, colors are approximately:  violet = 1.7 $\leq$ (\emph{M/M$_{\odot}$}) $<$ 2.0,(A \emph{SpTy's}); blue = 1.4 $\leq$ (\emph{M/M$_{\odot}$}) $<$ 1.7 (F \emph{SpTy's}); green = 0.9 $\leq$ (\emph{M/M$_{\odot}$}) $<$ 1.4 (G - mid-K \emph{SpTy's}); yellow = 0.6 $\leq$ (\emph{M/M$_{\odot}$}) $<$ 0.9 (late-K - early-M \emph{SpTy's}); orange = 0.3 $\leq$ (\emph{M/M$_{\odot}$}) $<$ 0.6 (mid-M \emph{SpTy's}); red = (\emph{M/M$_{\odot}$}) $<$ 0.3 ($>$M5 \emph{SpTy}).  The large cross at (X,Y,Z) = (0, 0, 0) represents the Sun.  The \emph{BPMG} shows extension in the \emph{X} direction, the direction toward the galactic center, a common feature among younger \emph{NYMGs} (T08).}
\label{bp_xyz}
\end{figure}

\clearpage

\begin{figure}
\epsscale{1.0}
\plotone{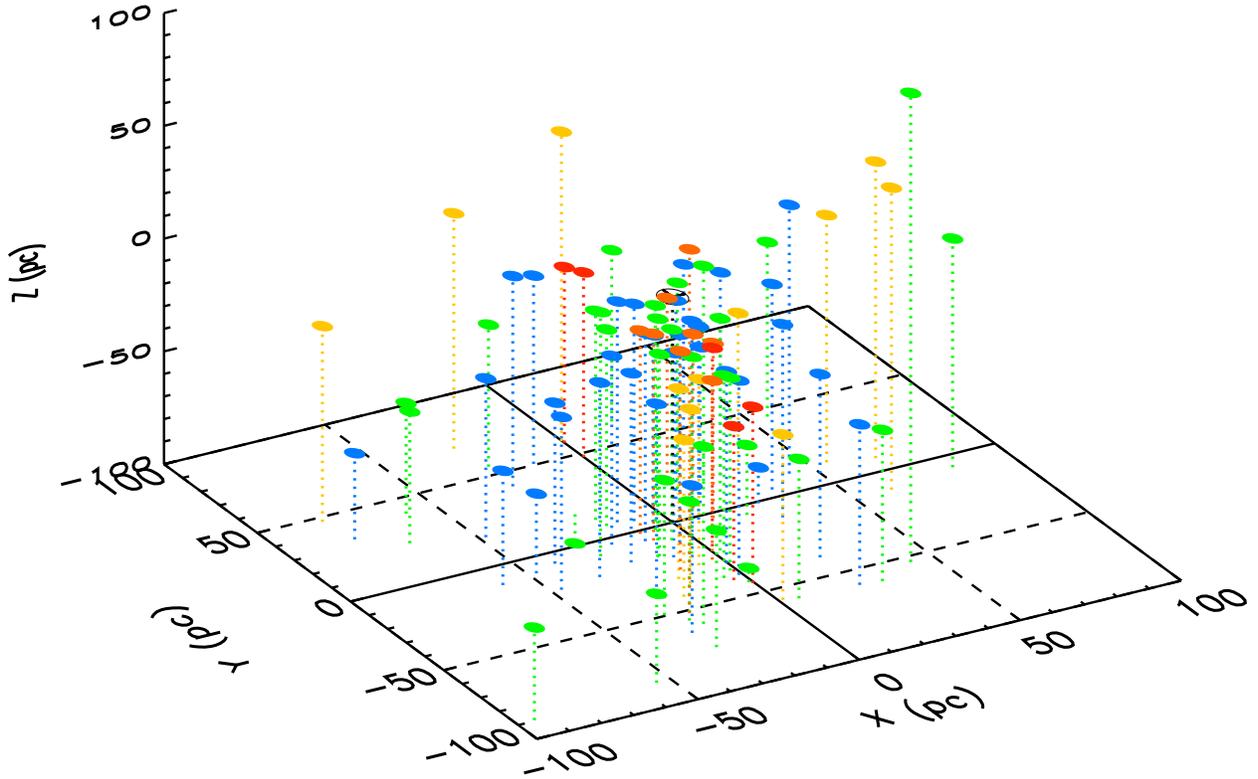}
\caption{\emph{ABDMG} \emph{XYZ} distribution, symbols are identical to Fig.~\ref{bp_xyz}.  The \emph{ABDMG} has a roughly uniform distribution and lacks the \emph{X} direction extension of the younger \emph{BPMG}, this feature may be due to its older age.  The lowest mass stars are near the Sun, probably a selection effect due to their low luminosities.}
\label{abd_xyz}
\end{figure}

\clearpage

\begin{figure}
\epsscale{0.9}
\plotone{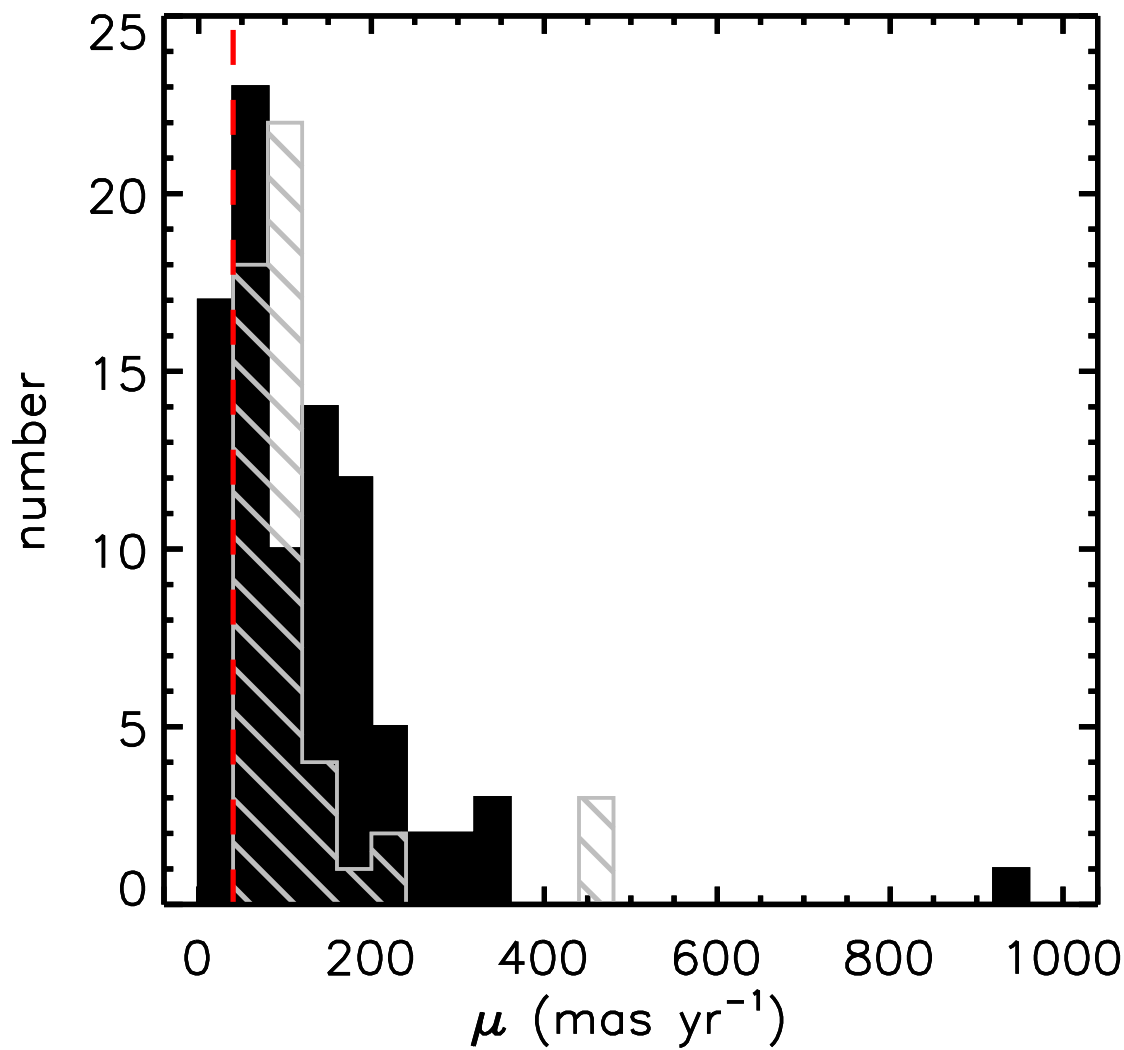}
\caption{Proper motion distributions of the known members of the \emph{BPMG} (gray, hashed) and \emph{ABDMG} (black, solid).  The larger spread in proper motions of \emph{ABDMG} members is in part due
to the larger space motion of the group, but may also be due to a larger spatial extension of the group due to its older age.  The dashed red line at $\mu$ = 40 mas yr$^{-1}$ represents the lower proper motion limit in our candidate search.  Most known members in the northern hemisphere have $\mu$ $\ge$ 40 mas yr$^{-1}$.  A color version of this figure is available in the electronic journal.}
\label{pm_plot}
\end{figure}

\clearpage

\begin{figure}
\epsscale{0.9}
\plotone{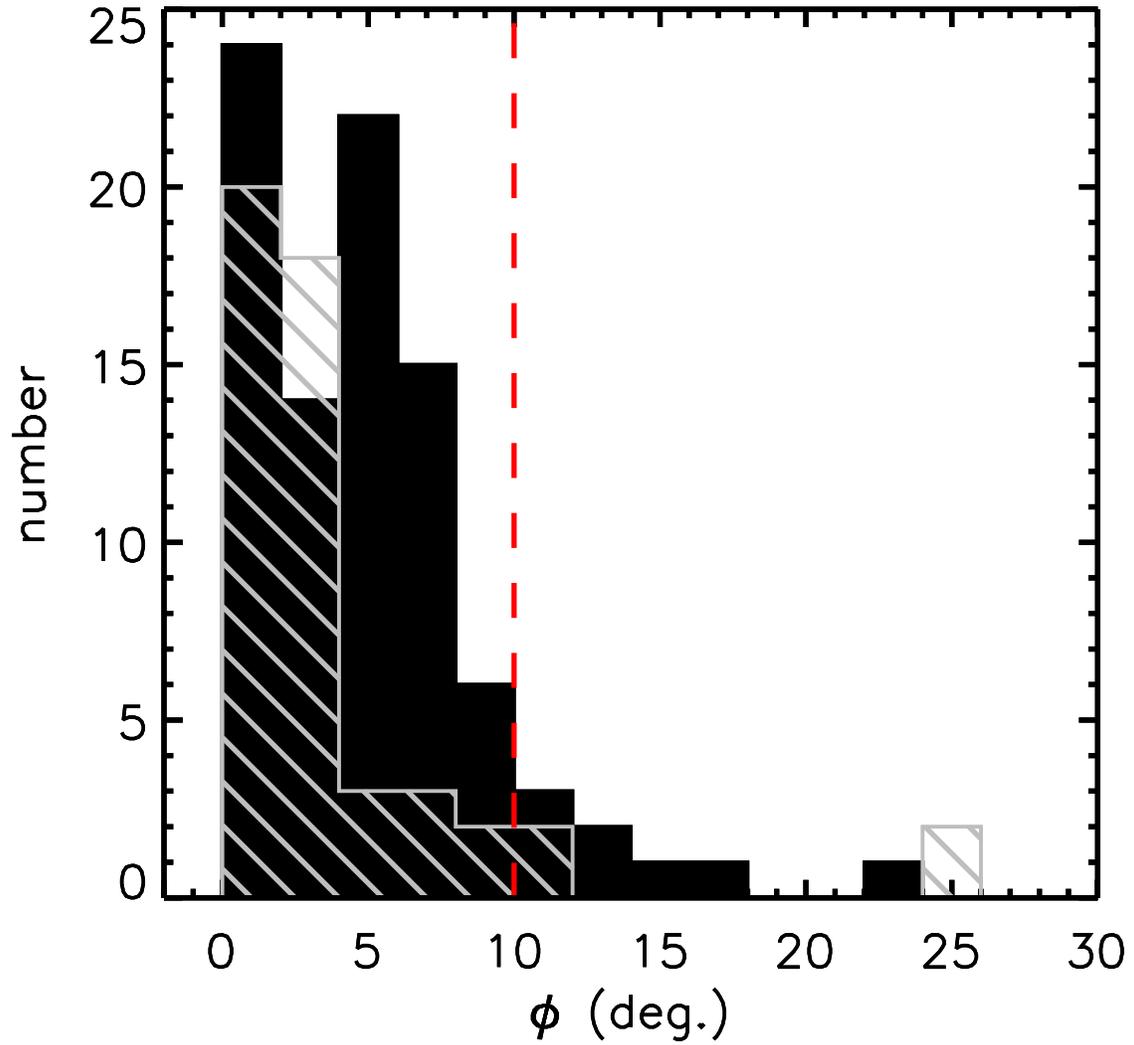}
\caption{$\phi$ distributions of the known members of the \emph{BPMG} and \emph{ABDMG}.  Color designations are the same as Fig.~\ref{pm_plot}.  The red dashed line at $\phi$ = 10$^{\circ}$ is the upper $\phi$ limit of our search.  A color version of this figure is available in the electronic journal.}
\label{phi_plot}
\end{figure}

\clearpage

\begin{figure}
\epsscale{0.9}
\plotone{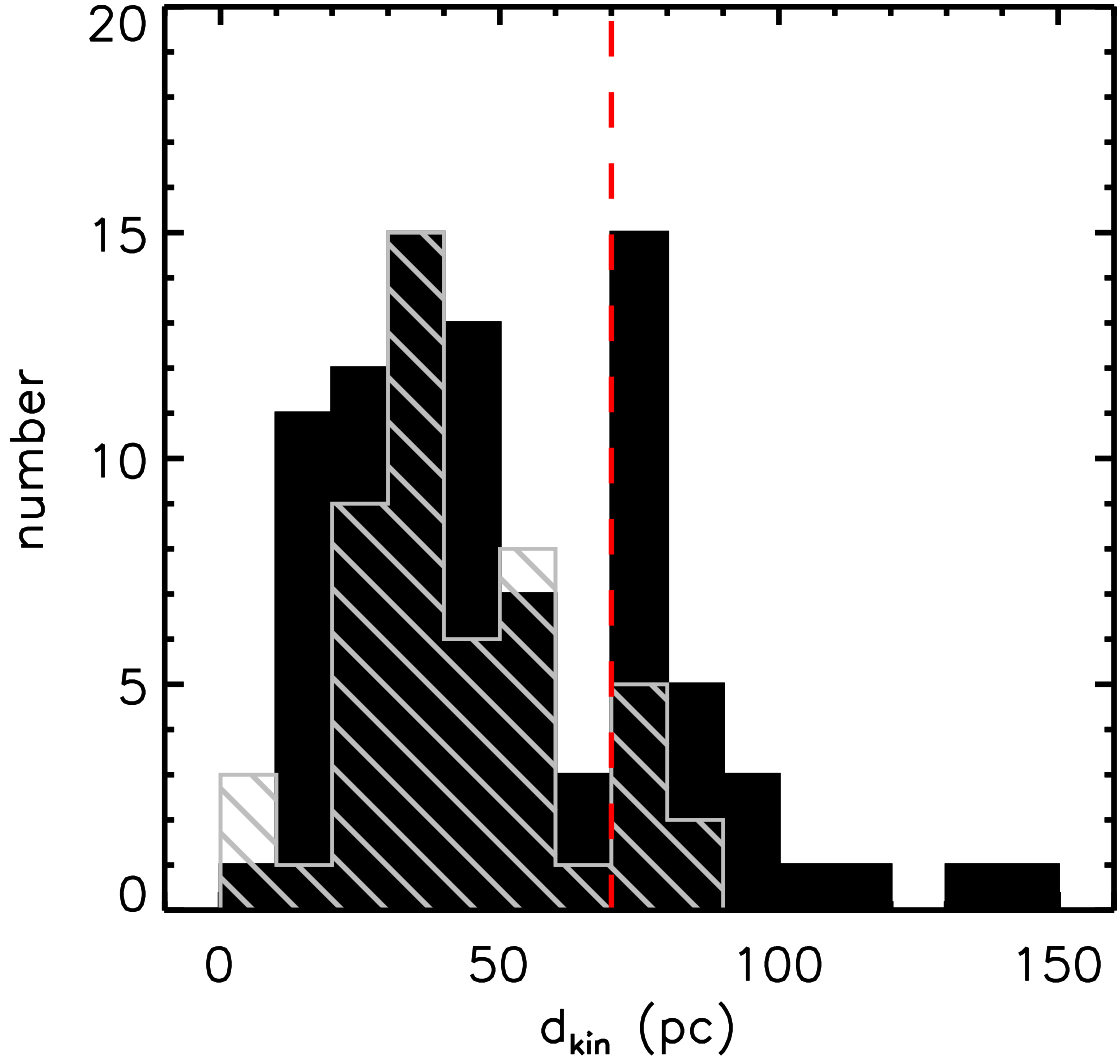}
\caption{\emph{d$_{kin}$} distributions of the known members of the \emph{BPMG} and \emph{ABDMG}.  Color designations are the same as Fig.~\ref{pm_plot}  Since known \emph{NYMG} members are plotted, \emph{d$_{kin}$} is the true distance.  We choose 70 pc as the upper \emph{d$_{kin}$} limit of our candidate search (dashed red line).  A color version of this figure is available in the electronic journal.}
\label{dkin_plot}
\end{figure}

\clearpage

\begin{figure}
\epsscale{0.70}
\plotone{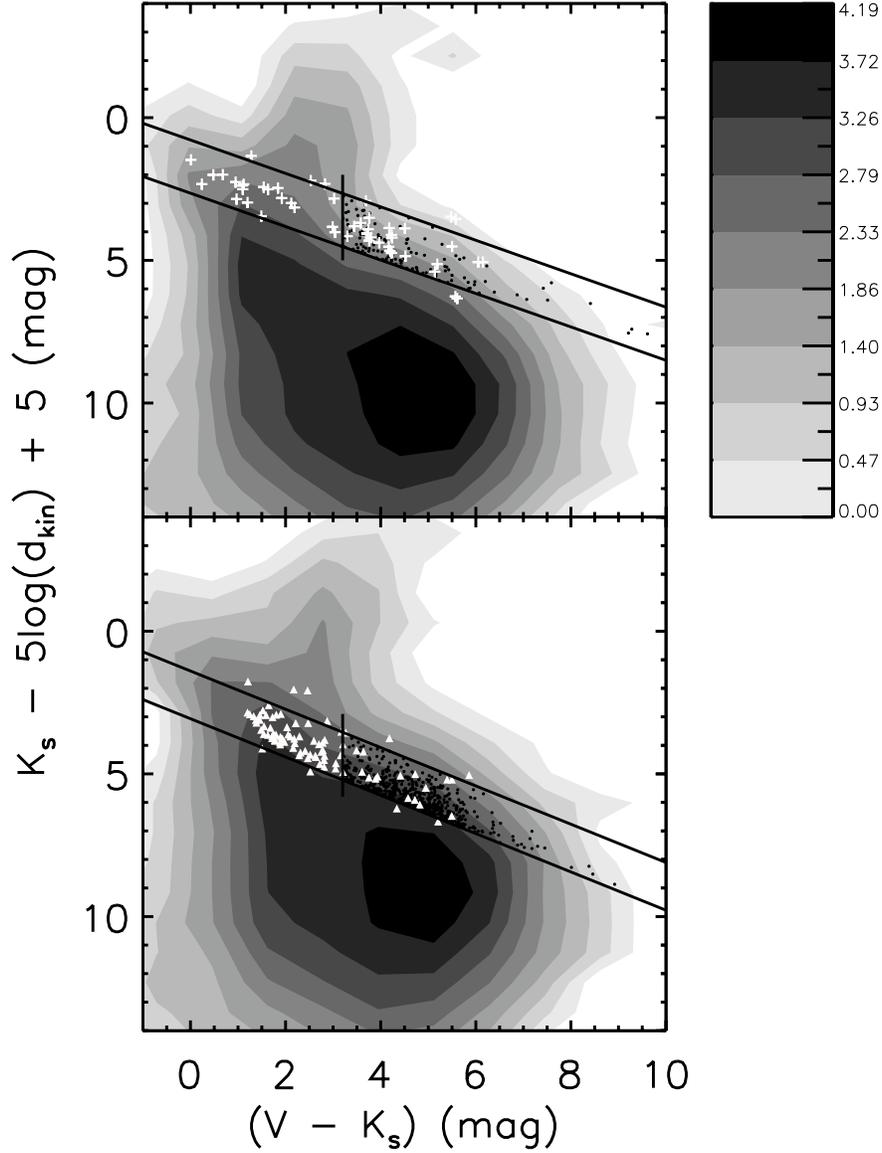}
\caption{Color-magnitude diagrams for \emph{BPMG} (top) and \emph{ABDMG} (bottom) candidates in the \emph{Tycho-2}, \emph{LSPM}, and \emph{SBK} catalogs.  \emph{M$_K$} was calculated using candidate \emph{d$_{kin}$}.  Shaded contours represent stars whose proper motions are consistent with group membership.  The shading is defined by the logarithm of the number density of these stars in \emph{M$_{K}$} \emph{vs.} (\emph{V-K$_s$}) space.  The logarithmic scale is given by the vertical bar to the right of the top plot.  White crosses and solid triangles represent known members of the respective groups.  The parallel black lines indicate a 2$\sigma$ uncertainty in a least-squares linear fit to the \emph{NYMG} cluster sequence, any star falling outside of the lines is rejected from the sample.  The vertical bar at (\emph{V-K$_{s}$}) = 3.2 indicates the color cut imposed in order to select stars later than $\sim$K5 and avoid the inclusion of giants.  Candidates selected by the search algorithm are represented by black dots, these are stars surviving all cuts imposed in the search, including \emph{d$_{kin}$} $\le$ 70 pc.}
\label{cmd_plot}
\end{figure}

\clearpage

\begin{figure}
\epsscale{0.9}
\plotone{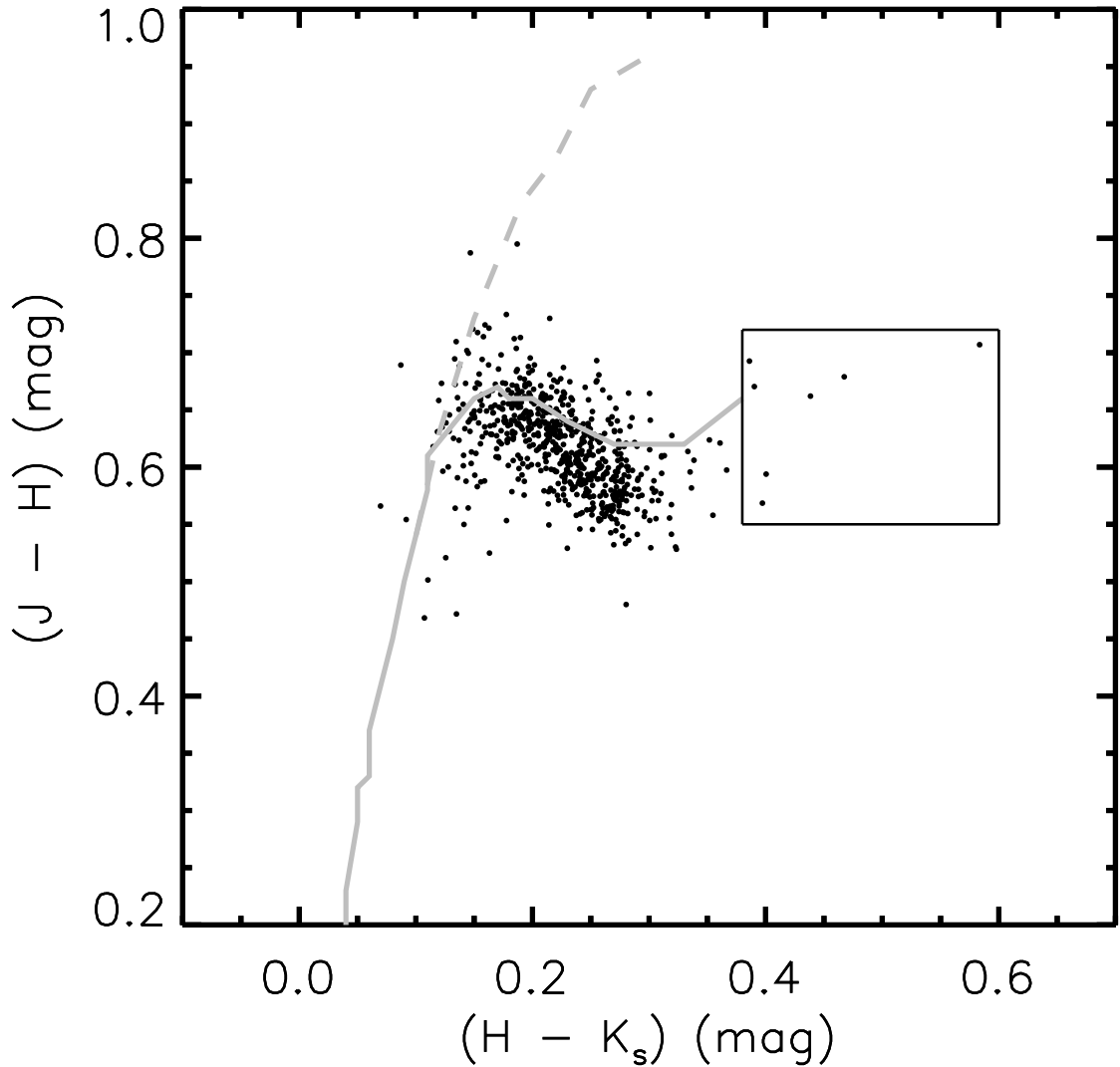}
\caption{Near IR color-color diagram of all candidates (black dots).  The expected main sequence (\emph{MS}) dwarf and late-type giant sequences are represented by the solid and dashed gray lines respectively.  As expected from the (\emph{V-K$_s$}) selection cut,  the vast majority of stars in the sample are late-type dwarfs.  Several outliers that appear to be giants or earlier-type dwarfs are removed from the sample.  The 7 candidates in the solid box have (\emph{H-K$_{s}$}) colors redder than \emph{MS} dwarfs, they are the ultracool candidates discussed in \S 6}     
\label{ccd_plot}
\end{figure}
 
 \clearpage
 
 \begin{figure}
\epsscale{0.9}
\plotone{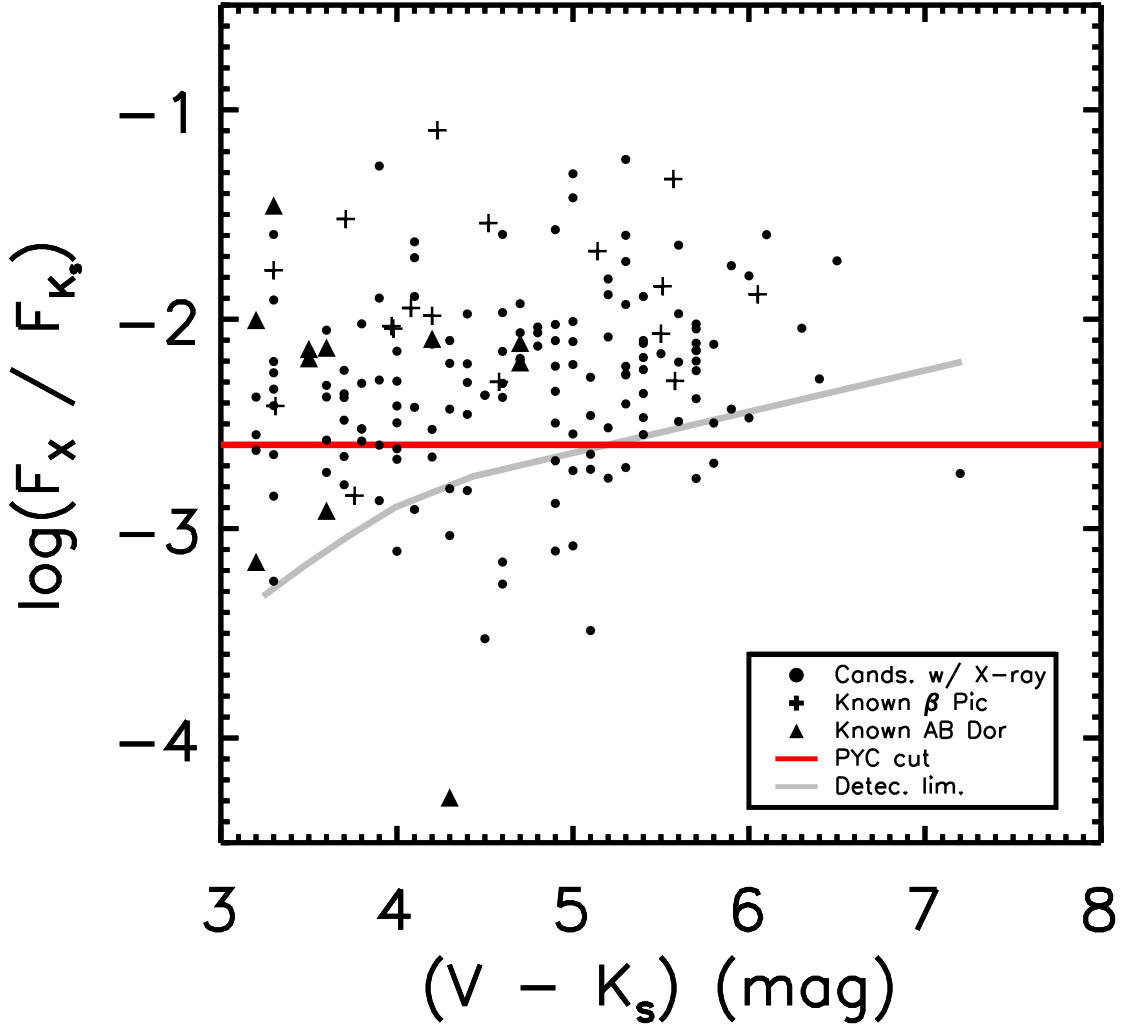}
\caption{X-ray flux ratios for candidates with counterparts in the \emph{ROSAT} catalogs.  Known late-type $\beta$ Pic and AB Dor members are plotted as crosses and upward triangles respectively.  Candidates with detected X-ray flux are represented with filled black circles.  Candidates lying above the red line at log(\emph{F$_X$/F$_{K_s}$}) = -2.6 have flux ratios consistent with youth as described in the text.  These stars are kept in the probable young candidate (\emph{PYC}) sample.  Candidates falling below this line are removed.  Known AB Dor member BD+01 2477 has a flux ratio well below the cut ($\sim$~-4.3) and should be investigated further.  The gray line is the estimated X-ray detection limit described in the text.  A color version of this figure is available in the electronic journal.}        
\label{xray_plot}
\end{figure}
 
 \clearpage

\begin{figure}
\epsscale{0.9}
\plotone{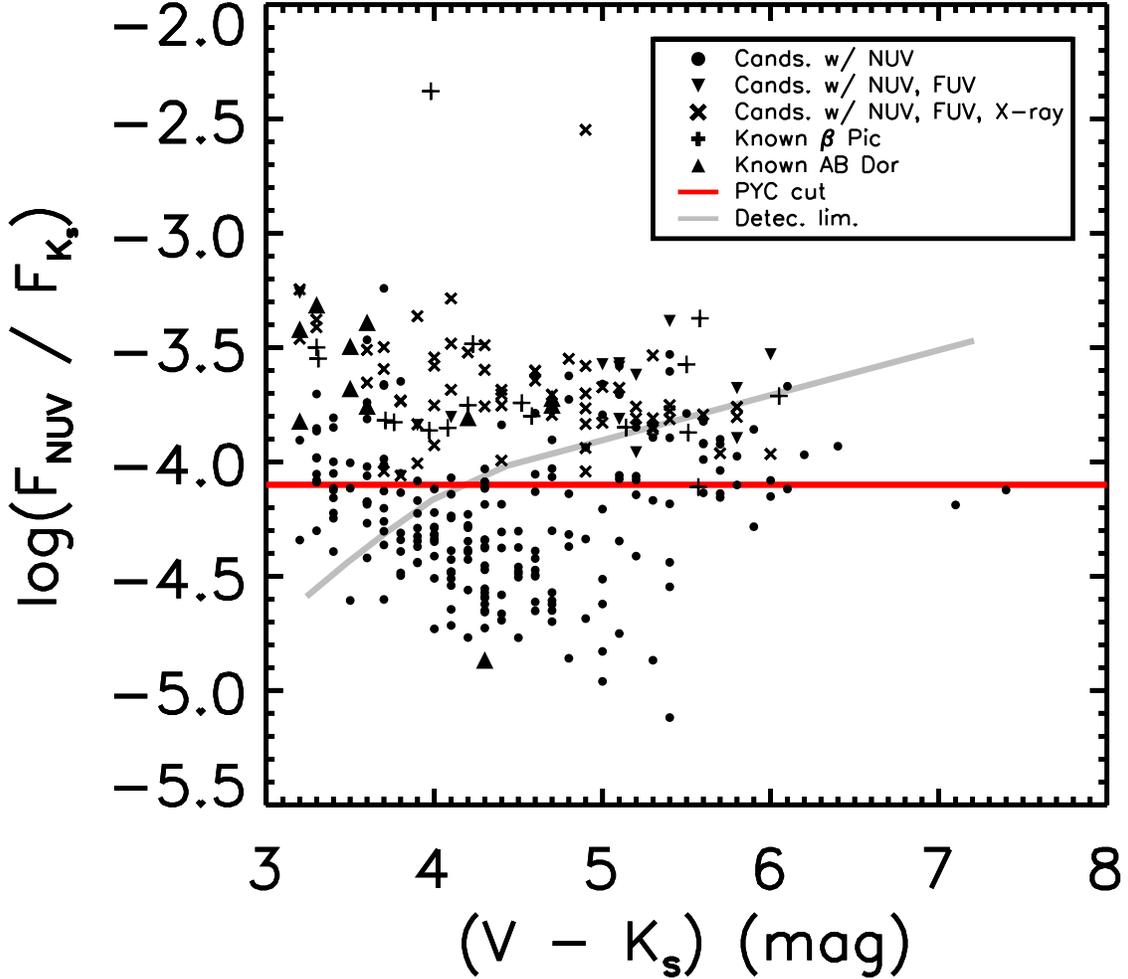}
\caption{\emph{NUV} flux ratios for candidates with counterparts in \emph{GALEX}.  Known \emph{NYMG} symbols are identical to Fig.~\ref{xray_plot}.  Candidates having only \emph{NUV} flux detections are represented by filled black circles.  The red line at log(\emph{F$_{NUV}$/F$_{K_s}$}) = -4.1 represents the cut imposed for youth in the text.  Candidates lying above this line are considered young and are kept in the probable young candidate (\emph{PYC}) sample.  Candidates falling below this line are removed.  Candidates with strong \emph{NUV}, \emph{FUV}, and X-ray are represented with a black X.  Downward facing triangles represent candidates with strong \emph{NUV} and \emph{FUV} but no X-ray that would have been missed if only \emph{ROSAT} data were considered.  One candidate not plotted was mostly likely observed during a flare and has \emph{NUV} and \emph{FUV} flux ratios -1.34 and -1.29 respectively.  BD+01 2477, a member of the AB Dor group, again has flux well below the youth cut ($\sim$~-4.8, see also R11).  The gray line is the estimated \emph{NUV} detection limit described in the text.  A color version of this figure is available in the electronic journal.}     
\label{uv_plot}
\end{figure}
 
 \clearpage

\begin{figure}
\epsscale{0.55}
\plotone{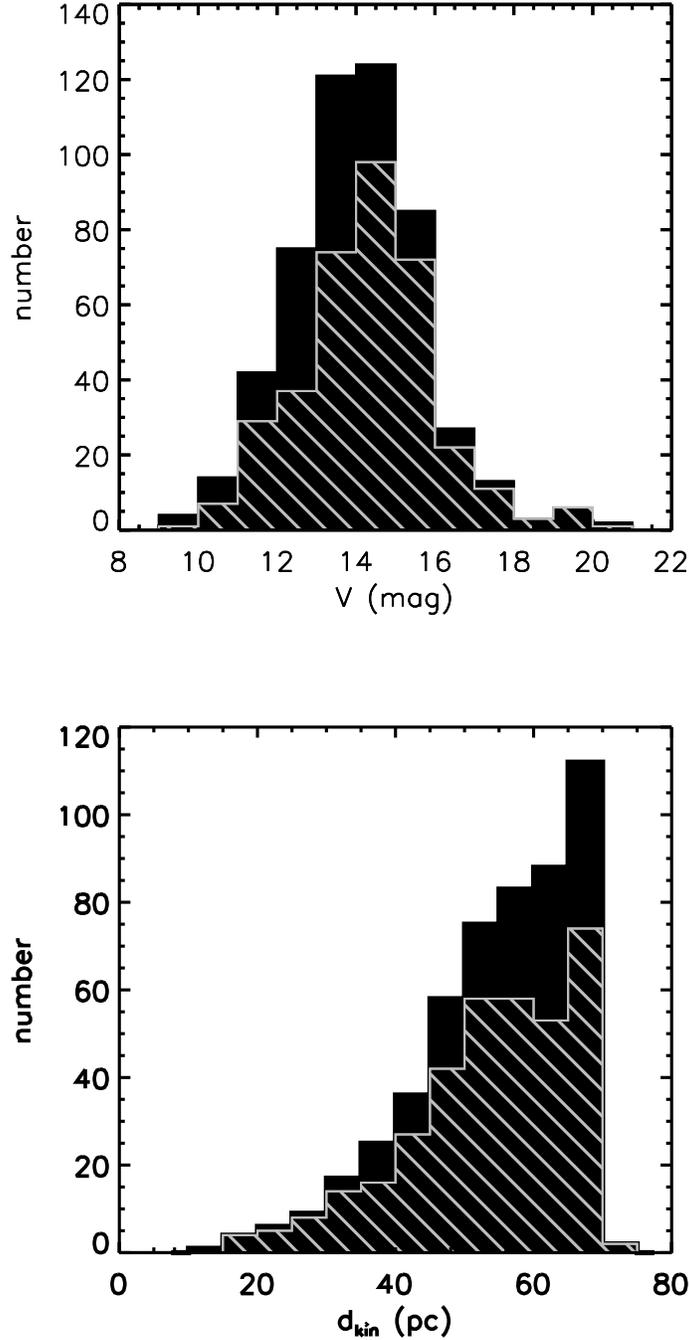}
\caption{\emph{V} (top) and \emph{d$_{kin}$} (bottom) distributions for candidates undetected (\emph{UCs}) by \emph{ROSAT} (solid, black) and \emph{GALEX} (hashed, gray).  The majority of \emph{UCs} have \emph{V}$\sim$13-15 and \emph{d$_{kin}$} $\gtrsim$50 pc.  This is consistent with most \emph{UCs} being true main-sequence interlopers with increasing numbers as search volume increases.  Fewer candidates are undetected by \emph{GALEX} since it is more sensitive than \emph{ROSAT}.  The sensitivity advantage of \emph{GALEX} is also apparent in the shift of the \emph{V} distribution to fainter magnitudes and in the approximate flattening of the \emph{d$_{kin}$} distribution after 50 pc.} 
\label{undetected_plot}
\end{figure}

 \clearpage

\begin{figure}
\epsscale{0.55}
\plotone{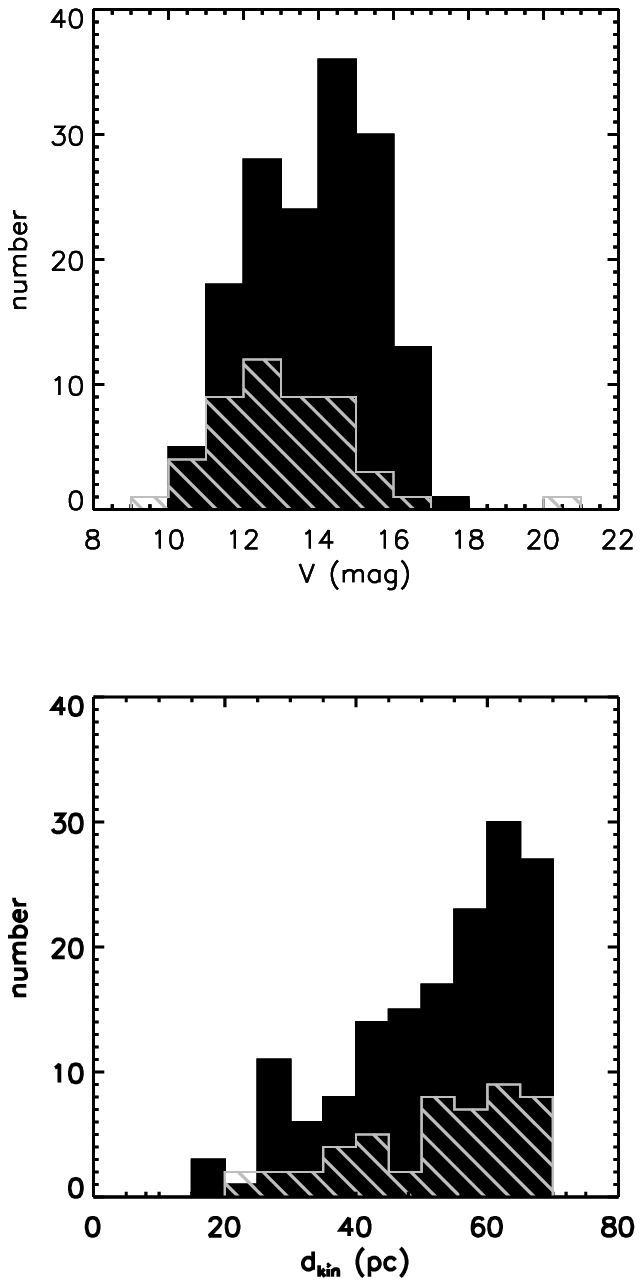}
\caption{Northern probable young candidate (\emph{PYC}) \emph{V} mag (top) and \emph{d$_{kin}$} (bottom) distributions.  \emph{BPMG} (hashed, gray) and \emph{ABDMG} (solid, black) histograms exhibit differences in both distributions.  \emph{BPMG} candidates are brighter on average and there are fewer at large predicted distances.  Because of its older age, the \emph{ABDMG} suffers from greater contamination as search volume increases.   The distributions indicate that \emph{BPMG} candidates are more promising targets for follow up and may yield a higher success rate.} 
\label{PYChist_plot}
\end{figure}

\clearpage

\begin{figure}
\epsscale{0.55}
\plotone{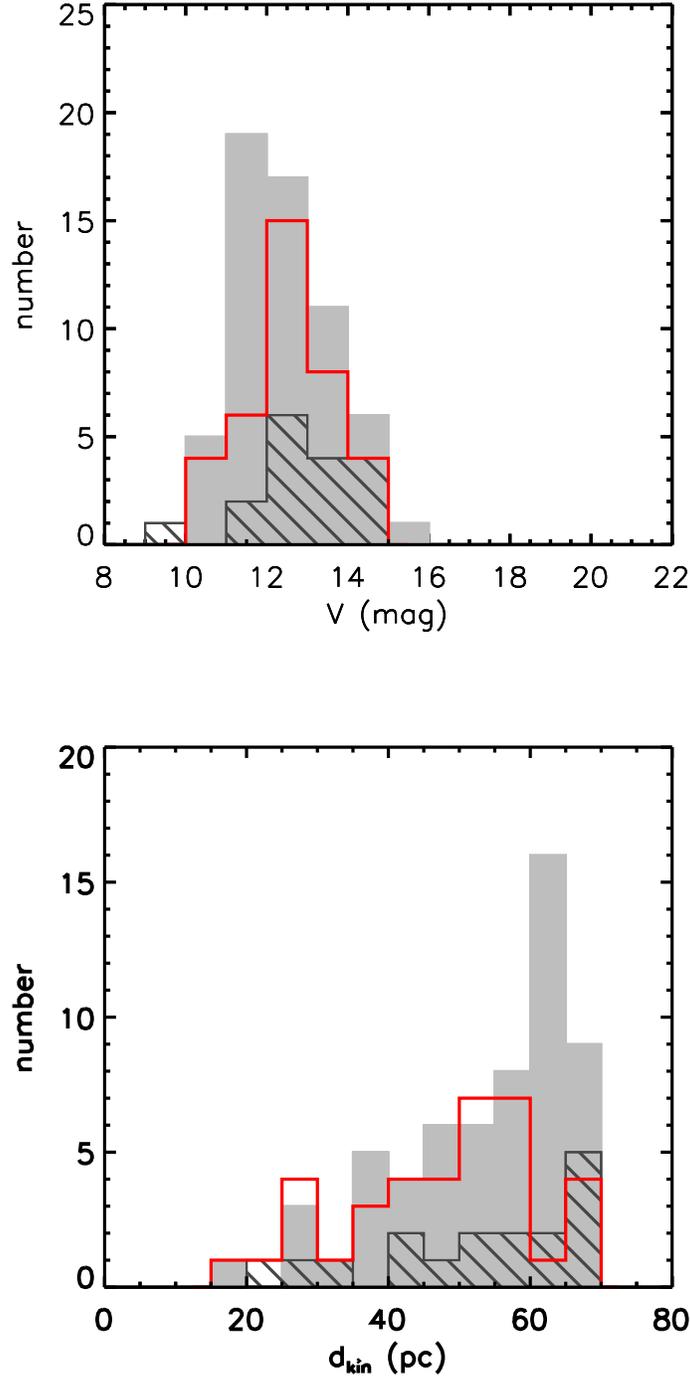}
\caption{Candidates with reliable youth (\emph{CWRY}) activity index ({\tt A}) \emph{V} mag (top) and \emph{d$_{kin}$} (bottom) distributions.  Histograms represent \emph{CWRYs} with $\Sigma${\tt A} = 1 (solid, gray),  2 (hashed, dark gray), and 3 (open, red). Many of the \emph{CWRYs} are candidates having only one strong activity indicator; typically only strong X-ray or \emph{NUV}.  There are more \emph{CWRYs} with $\Sigma${\tt A} = 3, those with the strongest indication of youthful activity, than $\Sigma${\tt A} = 2.  The $\Sigma${\tt A} = 3 distance distribution also peaks at smaller \emph{d$_{kin}$}.  This combination of multiple strong activity indicators, relatively bright \emph{V} mags, and smaller predicted distances makes these candidates high priority for follow up investigations.  A color version of this figure is available in the electronic journal.} 
\label{CWRY_plot}
\end{figure}

\clearpage

\begin{figure}
\epsscale{0.55}
\plotone{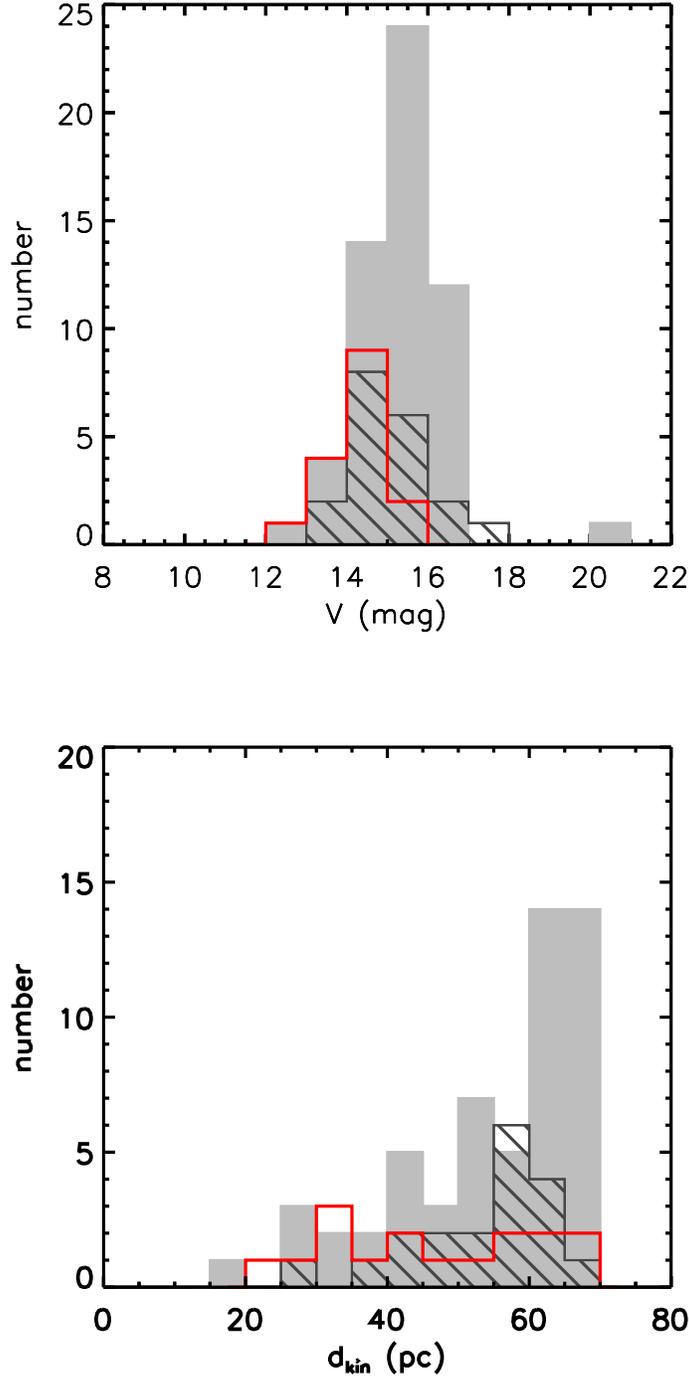}
\caption{Candidates with ambiguous youth (\emph{CWAY}) activity index ({\tt A}) \emph{V} mag (top) and \emph{d$_{kin}$} (bottom) distributions.  The three histograms represent \emph{CWAYs} with the same cumulative {\tt A} values described in Fig.~\ref{CWRY_plot}.  Because \emph{CWAYs} are candidates with redder (\emph{V-K$_s$}) colors the \emph{V} mag distributions are systematically shifted to fainter magnitudes.  The \emph{d$_{kin}$} distribution of the $\Sigma${\tt A} = 3 \emph{CWAY} subsample is approximately flat.  This feature is indication that contamination does not increase as volume increases in this subsample.  Thus these candidates are prime for follow up.  A color version of this figure is available in the electronic journal.} 
\label{CWAY_plot}
\end{figure}

\clearpage

\begin{figure}
\epsscale{0.55}
\plotone{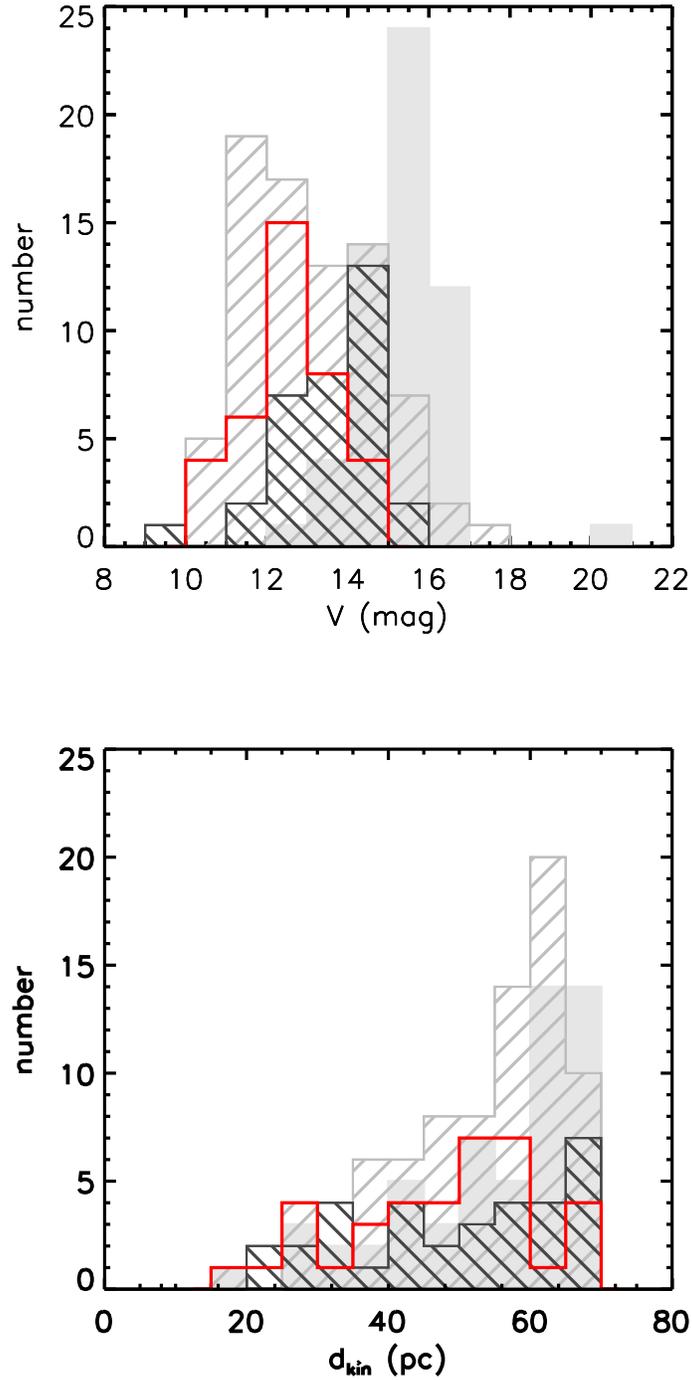}
\caption{Probable young candidate (\emph{PYC}) priority index ({\tt P}) \emph{V} mag (top) and \emph{d$_{kin}$} (bottom) distributions.  Histograms represent candidates with {\tt P} = 1 (solid,gray), 2 (hashed, light gray), 3 (hashed, dark gray), and 4 (open, red).  Candidates with {\tt P} = 4 exhibit many, reliable, strong activity indicators and are the highest priority for follow up.  Follow up priority decreases with {\tt P} index.  A color version of this figure is available in the electronic journal.}
\label{P_plot}
\end{figure}

\clearpage

\begin{figure}
\epsscale{0.95}
\plotone{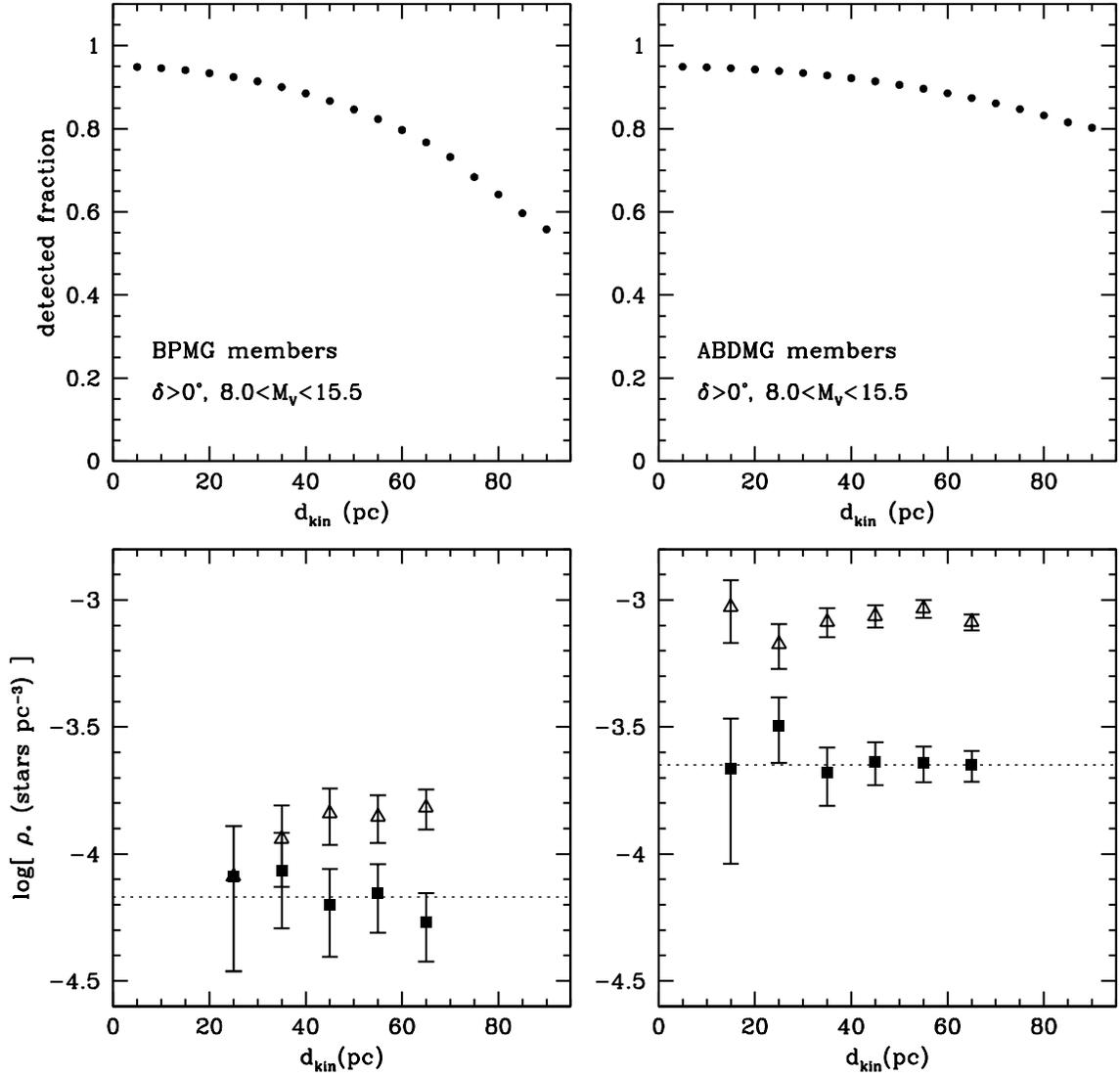}
\caption{Estimated completeness of the search and predicted number
   densities of northern, low-mass stars in each of the $\beta$ Pic and AB Dor
   moving groups. Upper panels: expected rate of detection for stars
   with absolute magnitudes $8.0<M_V<15.5$, typical for young M
   dwarfs. The main source of incompleteness comes from the proper
   motion limit ($\mu>40$ mas yr$^{-1}$) of the source
   catalog; the incompleteness increases with distance. Lower-panels:
   volume densities of $\beta$ Pic (left) and AB Dor (right) moving
   group candidates, corrected for incompleteness. Estimates based on
   the full candidate list (open triangles) are compared to estimates using
   only probable young candidates with strong activity (\emph{PYCs}, filled squares). The latter
   provide an upper limit estimate for the density of the group in the
   surveyed volume.}
\label{comp_plot}
\end{figure}

\clearpage








\end{document}